\documentclass{IEEEtran}
\usepackage{amsthm,amsmath,amssymb,dsfont,graphicx}
\usepackage{calc}

\theoremstyle{plain} \newtheorem{thm}{Theorem}
\newtheorem{prop}[thm]{Proposition} 
 \newtheorem{lem}[thm]{Lemma}
\newtheorem{coro}[thm]{Corollary} \theoremstyle{definition}
\newtheorem{defn}[thm]{Definition} \newtheorem{rem}[thm]{Remark}

\newcommand{\cE}{\mathcal{E}}
\newcommand{\cG}{\mathcal{G}}
\newcommand{\good}{\mathcal{W}}
\newcommand{\cB}{\mathcal{B}}
\newcommand{\eps}{\epsilon}
\newcommand{\bbE}{\mathbb{E}}
\newcommand{\piu}{\pi^{(u)}}

\title{Graph-Constrained Group Testing}
\author{Mahdi Cheraghchi,  \IEEEmembership{Member, IEEE},
Amin Karbasi,  \IEEEmembership{Student Member, IEEE}, \\
Soheil~Mohajer, \IEEEmembership{Member, IEEE},
and Venkatesh Saligrama,  \IEEEmembership{Senior Member, IEEE}

  \thanks{{\footnotesize 
      Mahdi Cheraghchi is with the Department of Computer Science,
      University of Texas at Austin, Austin, TX 78701, USA (email: mahdi@cs.utexas.edu).
     Amin Karbasi is with the School of Computer and Communication
      Sciences, Ecole Polytechnique F\'{e}d\'{e}rale de Lausanne
      (EPFL), CH-1015 Lausanne, Switzerland (email: amin.karbasi@epfl.ch).
      Soheil Mohajer is with the Department
      of Electrical Engineering, Princeton University, Princeton, NJ 08544, USA
      (email: smohajer@princeton.edu).
      Venkatesh Saligrama is
      with the Department of Electrical and Computer Engineering at
      Boston University, Boston, MA 02215, USA (email: srv@bu.edu).
      Part of research was done while M.~Cheraghchi and S.~Mohajer were
      with the School of Computer and Communication Sciences, EPFL, Switzerland.
      M.~Cheraghchi was supported
      by the ERC Advanced investigator grant 228021 of A.~Shokrollahi.
      S.~Mohajer was supported by ERC Starting Investigator grant \#240317.
      V.~Saligrama was supported by the U.S. Department of Homeland
      Security under Award Number 2008-ST-061-ED0001, NSF CPS Award 0932114, NSF CAREER
      Award Number ECS 0449194.  The views and conclusions contained
      in this document are those of the authors and should not be
      interpreted as necessarily representing the official policies,
      either expressed or implied, of the U.S. Department of Homeland
      Security or the US National Science Foundation. A preliminary
      summary of this work appeared (under the same title) in
      proceedings of the 2010 IEEE International Symposium on
      Information Theory. }}  }

\begin{document}

\maketitle

\maketitle

\begin{abstract}
  Non-adaptive group testing involves grouping arbitrary subsets of
  $n$ items into different pools. Each pool is then tested and
  defective items are identified. A fundamental question involves
  minimizing the number of pools required to identify at most $d$
  defective items.  Motivated by applications in network tomography,
  sensor networks and infection propagation, a variation of group
  testing problems on graphs is formulated. Unlike conventional group
  testing problems, each group here must conform to the constraints
  imposed by a graph. For instance, items can be associated with
  vertices and each pool is any set of nodes that must be path
  connected. In this paper, a test is associated with a random walk. In
  this context, conventional group testing corresponds to the special
  case of a complete graph on $n$ vertices.

  For interesting classes of graphs a rather surprising result is
  obtained, namely, that the number of tests required to identify $d$
  defective items is substantially similar to what is required in
  conventional group testing problems, where no such constraints on
  pooling is imposed. Specifically, if $T(n)$ corresponds to the
  mixing time of the graph $G$, it is shown that with
  $m=O(d^2T^2(n)\log(n/d))$ non-adaptive tests, one can identify the
  defective items. Consequently, for the Erd\H{o}s-R\'enyi random
  graph $G(n,p)$, as well as expander graphs with constant spectral
  gap, it follows that $m=O(d^2\log^3n)$ non-adaptive tests are
  sufficient to identify $d$ defective items. Next, a specific
  scenario is considered that arises in network tomography, for which
  it is shown that $m=O(d^3\log^3n)$ non-adaptive tests are sufficient
  to identify $d$ defective items. Noisy counterparts of the graph
  constrained group testing problem are considered, for which parallel
  results are developed. We also briefly discuss extensions to compressive sensing on graphs.

\end{abstract}


\section{Introduction}
\IEEEPARstart{I}{n} this paper we introduce the graph constrained group testing problem
motivated by applications in network tomography, sensor networks and
infection propagation. While group testing theory (see ~\cite{ref:Dor43,du} and
more recently~\cite{atia09}), and its numerous applications, such as
industrial quality assurance~\cite{ref:SG59}, DNA library
screening~\cite{ref:PL94}, software testing~\cite{ref:BG02}, and
multi-access communications~\cite{ref:Wol85}, have been systematically
explored, the graph constrained group testing problem is new to the
best of our knowledge.

Group testing involves identifying at most $d$ defective items out of
a set of $n$ items. In non-adaptive group testing, which is the
subject of this paper, we are given an $m \times n$ binary matrix,
$M$, usually referred to as a test or measurement matrix. Ones on the
$j$th row of $M$ indicate which subset of the $n$ items belongs to the
$j$th pool. A test is conducted on each pool; a positive outcome
indicating that at least one defective item is part of the pool; and a negative
test indicating that no defective items are part of the pool. The
conventional group testing problem is to design a matrix $M$ with
minimum number of rows $m$ that guarantees error free identification
of the defective items. While the best known (probabilistic) pooling
design requires a test matrix with $m = O(d^2 \log (n/d))$ rows, and
an almost-matching lower bound of $m = \Omega(d^2 (\log n)/(\log d))$
is known on the number of pools (cf.\ \cite[Chapter 7]{du}), the size
of the optimal test still remains open.

Note that in the standard group testing problem the test matrix $M$
can be designed arbitrarily. In this paper we consider a
generalization of the group testing problem to the case where the
matrix $M$ must conform to constraints imposed by a graph
$G=(V,E)$. In general, as we will describe shortly, such problems
naturally arise in several applications such as network
tomography~\cite{Duffield,NgTh}, sensor networks~\cite{THSNET}, and
infection propagation~\cite{CHKV:09}. While the graph constrained group testing
problem has been alluded to in these applications, the problem of test
design or the characterization of the minimum number of tests, to the
best of our knowledge, has not been addressed before. In this light our
paper is the first to formalize the graph constrained group testing
problem. In our graph group testing problem the $n$ items are either
vertices or links (edges) of the graph; at most $d$ of them are
defective. The task is to identify the defective vertices or
edges. The test matrix $M$ is constrained as follows: for items
associated with vertices each row must correspond to a subset of
vertices that are connected by a path on the graph; similarly, for items
associated with links each row must correspond to links that
form a path on $G$. The task is to design an $m \times
n$ binary test matrix with minimum number of rows $m$ that guarantees
error free identification of the defective items.

We will next describe several applications, which illustrate the graph
constrained group testing problem.

\subsection{Network Tomography \& Compressed Sensing over Graphs}
For a given network, identification of congested links from end-to-end
path measurements is one of the key problems in network tomography
\cite{NgTh}, \cite{Duffield}. In many settings of today's IP networks,
there is one or a few links along the path which cause the packet
losses in the path. Finding the locations of such congested links is
sufficient for most of the practical applications.

\begin{figure}[t]
   \centering
   \includegraphics[width=8cm]{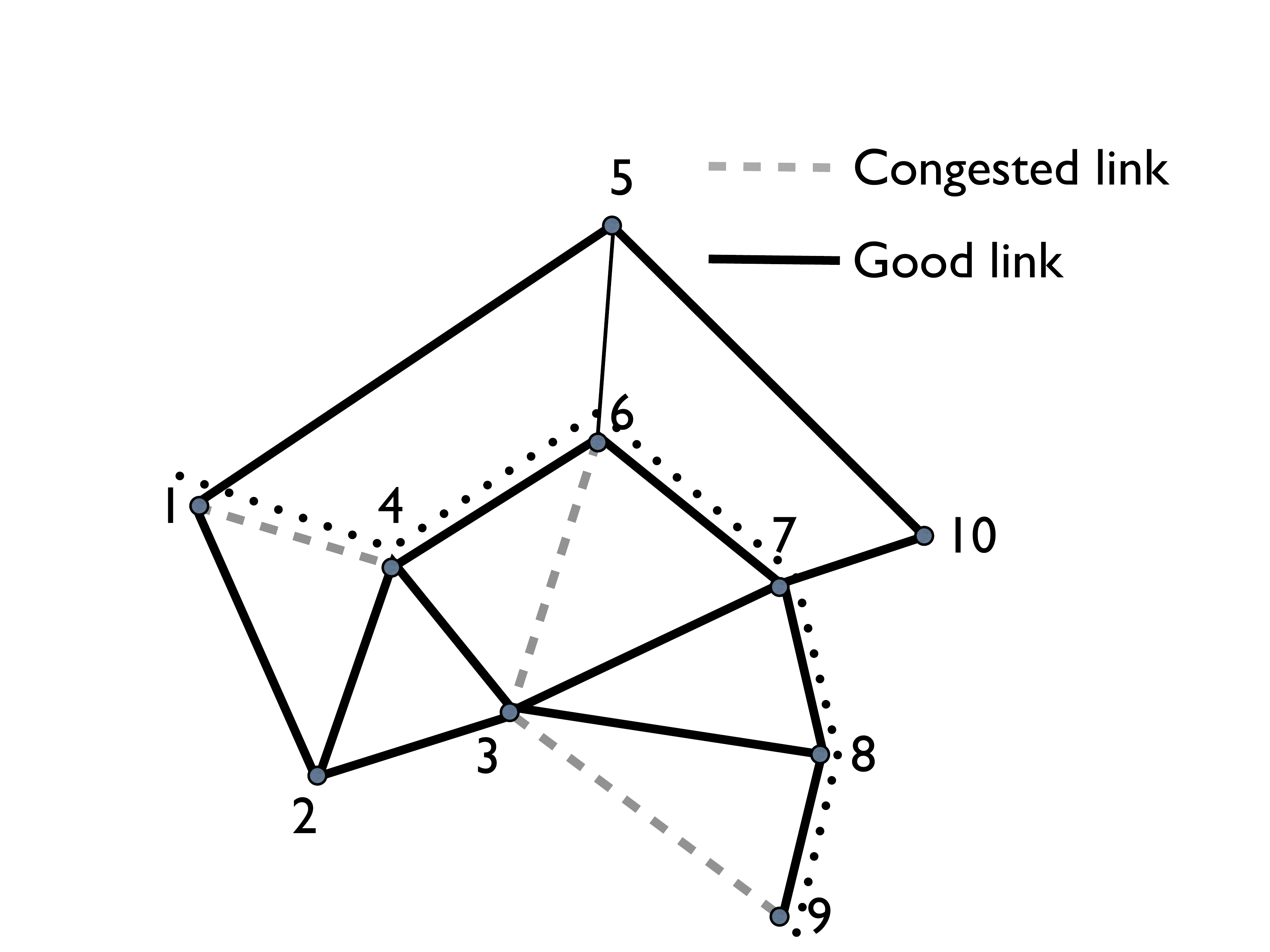}
   \caption{The route $1\rightarrow4\rightarrow 6 \rightarrow 7
     \rightarrow 8\rightarrow 9$ is valid while $2\rightarrow
     6\rightarrow 5$ is not. }\label{fig:network}
 \end{figure}

This problem can be understood as a graph-constrained group testing as
follows. We model the network as a graph $G=(V,E)$ where the set $V$
denotes the network routers/hosts and the set $E$ denotes the
communication links (see Fig.~\ref{fig:network}). Suppose, we have a
monitoring system that consists of one or more end hosts (so called
\textit{vantage} points) that can send and receive packets. Each
vantage point sends packets through the network by assigning the
routes and the end hosts.

All measurement results (i.e., whether each packet has reached its
destination) will be reported to a central server whose
responsibility is to identify the congested links. Since the network
is given, not any route is a valid one. A vantage point can only
assign those routes which form a path in the graph $G$.  The question
of interest is to determine the number of measurements that is needed
in order to identify the congested links in a given network.

We primarily deal with Boolean operations on binary valued variables in this paper, namely, link states are binary valued and the measurements are boolean operations on the link states. Nevertheless, the techniques described here can be extended to include non-boolean operations and non-binary variables as well. Specifically, suppose there are a sparse set of links that take on non-zero values. These non-zero values could correspond to packet delays, and packet loss probabilities along each link. Measurements along each path provides aggregate delay or aggregate loss along the path. The set of paths generated by $m$ random walks forms a $m \times |E|$ routing matrix $M$. For an appropriate choice of $m$ and graphs studied in this paper, it turns out (see \cite{weiyu}) that such routing matrices belongs to the class of so called expander matrices. These expander type properties in turn obey a suitable type of restricted-isometry-property (1-RIP) \cite{berinde}. Such properties in turn are sufficient for recovering sparse vectors using $\ell_1$ optimization techniques. Consequently, the results of this paper have implications for compressed sensing on graphs.

\subsection{Sensor Networks}
The network tomography problem is further compounded in wireless
sensor networks (WSN). As described in \cite{THSNET} the routing
topology in WSN is constantly changing due to the inherent ad-hoc
nature of the communication protocols. The sensor network is static
with a given graph topology such as a geometric random graph. Sensor
networks can be monitored passively or actively. In passive
monitoring, at any instant, sensor nodes form a tree to route packets
to the sink. The routing tree constantly changes unpredictably but
must be consistent with the underlying network connectivity. A test is
considered positive if the arrival time is significantly large, which
indicates that there is at least one defective sensor node or a congested
link. The goal is to identify defective links or sensor nodes based on
packet arrival times at the sink.  In active monitoring network nodes
continuously calculate some high level, summarized information such as
the average or maximum energy level among all nodes in the
network. When the high level information indicates congested links, a
low level and more energy consuming procedure is used to accurately
locate the trouble spots.

\subsection{Infection Propagation}
\begin{figure}[t]
  \centering
  \includegraphics[width=\columnwidth]{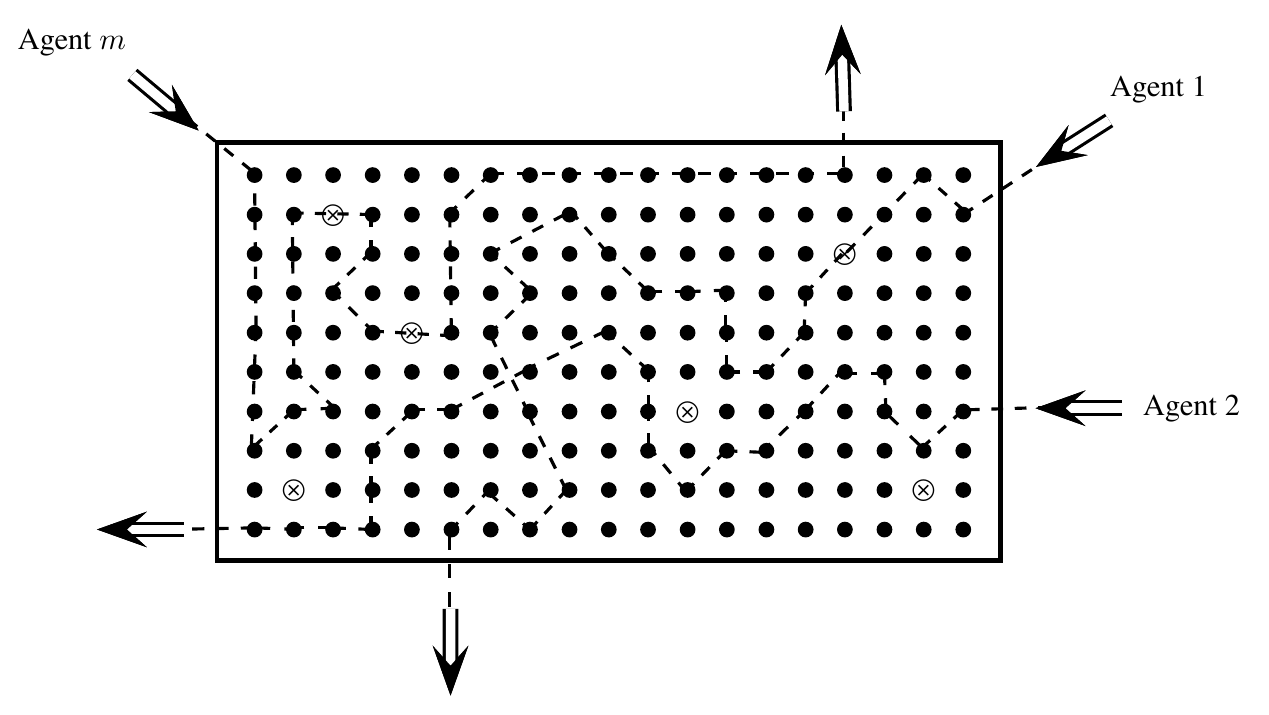}
  \caption{Collective sampling using agents. The~$\otimes$ symbols
    represent infected people whereas the healthy population is
    indicated by~$\bullet$~symbols. The dashed lines show the group of
    people contacted by each agent~\cite{CHKV:09}. }\label{fig:agents}
\end{figure}

Suppose that we have a large population where only a small number of
people are infected by a certain viral sickness (e.g., a flu
epidemic). The task is to identify the set of infected individuals by
sending agents among them. Each agent contacts a pre-determined or
randomly chosen set of people. Once an agent has made contact with an
infected person, there is a chance that he gets infected, too. By the
end of the testing procedure, all agents are gathered and tested for
the disease.  While this problem has been described in \cite{CHKV:09},
the analysis ignores the inherent graph constraints that need to be
further imposed. It is realistic to assume that, once an agent has
contacted a person, the next contact will be with someone in close
proximity of that person. Therefore, in this model we are given a
random geometric graph that indicates which set of contacts can be
made by an agent (see Fig.~\ref{fig:agents}). Now, the question is to
determine the number of agents that is needed in order to identify the
set of infected people.

These applications present different cases where graph constrained
group testing can arise. However, there are important distinctions.  In
the wired network tomography scenario the links are the items and each
row of the matrix $M$ is associated with a route between any two
vantage points. A test is positive if a path is congested, namely, if
it contains at least one congested link. Note that in this case since
the routing table is assumed to be static, the route between any two
vantage points is fixed. Consequently, the matrix $M$ is deterministic
and the problem reduces to determining whether or not the matrix $M$
satisfies identifiability.

Our problem is closer in spirit to the wireless sensor network
scenario. In the passive case the links are the items and each row of
the matrix $M$ is associated with a route between a sensor node and
the sink. A test is positive if a path is congested, namely, if it
contains at least one congested link. Note that in this case since the
routing table is constantly changing, the route between a sensor node
and the sink is constantly changing as well. Nevertheless the set of
possible routes must be drawn from the underlying connectivity
graph. Consequently, the matrix $M$ can be assumed to be random and
the problem is to determine how many different tests are required to
identify the congested links. Note that, in contrast to the wired
scenario, tests conducted between the same sensor node and sink yields
new information here. A similar situation arises in the active
monitoring case as well. Here one could randomly query along different
routes to determine whether or not a path is congested. These tests
can be collated to identify congested links. Note that in the active
case the test matrix $M$ is amenable to design in that one could
selectively choose certain paths over others by considering weighted
graphs.

Motivated by the WSN scenario we describe pool designs based on random
walks on graphs. As is well known a random walk is the state evolution
on a finite reversible Markov chain. Each row of the binary test
matrix is derived from the evolution of the random walk, namely, the
ones on the $j$th row of $M$ correspond to the vertices visited by the
$j$th walk. This is close to the WSN scenario because as in the WSN
scenario the path between two given nodes changes randomly. We develop
several results in this context.

First, we consider random walks that start either at a random node or
an arbitrary node but terminate after some appropriately chosen
number of steps $t$.
By optimizing the length of the walk we arrive at an interesting
result for important classes of graphs. Specifically we show that
the number of tests required to identify $d$ defective items is
substantially similar to that required in conventional group testing
problems, except the fact that an extra term appears which captures the
topology of the underlying graph. The best
known result for the number of tests required when no graphical
constraints are imposed scales as $O(d^2\log(n/d))$. For the graph
constrained case we show that with $m=O(d^2T^2(n)\log(n/d))$ non-adaptive
tests one can identify the defective items, where $T(n)$ corresponds
to the mixing time of the underlying graph $G$. Consequently, for the
Erd\H{o}s-R\'enyi random graph $G(n,p)$ with $p =\Omega( (\log^2 n)/n)$, as well as expander graphs
with constant spectral gap, it follows that $m=O(d^2\log^3n)$
non-adaptive tests are sufficient to identify $d$ defective items.
In particular, for a complete graph where no pooling constraint is imposed,
we have $T(n)=1$, and therefore, our result subsumes the well-known result
for the conventional group testing problem.

Next we consider unbounded-length random walks that originate at a
source node and terminate at a sink node. Both the source node and
the sink node can either be arbitrary or be chosen uniformly at
random. This directly corresponds to the network tomography problem
that arises in the WSN context. This is because the source nodes can
be viewed as sensor nodes, while the sink node maybe viewed as the
fusion center, where data is aggregated. At any instant, we can assume
that a random tree originating at the sensor nodes and terminating at
the sink is realized. While this random tree does not have cycles,
there exist close connections between random walks and randomly
generated trees. Indeed, it is well known that the so called loop-erased random walks, obtained by systematically erasing loops in random walks, to obtain spanning trees, is a method for sampling spanning trees from a uniform distribution~\cite{wilson}. In this scenario, we show
that $m=O(d^3\log^3n)$ non-adaptive tests are sufficient to identify
$d$ defective items. By considering complete graphs we also establish
that the cubic dependence on $d$ in this result cannot be improved.

We will also consider noisy counterparts of the graph constrained
group testing problem, where the outcome of each measurement
may be independently corrupted (flipped) with probability\footnote{It
is clear that if $q>1/2$, one can first flip all the outcomes, and
then reduce the problem to the $q<1/2$ regime. For $q=1/2$, 
since we only observe purely random noise, there is no hope to recover
from the errors.} $0\leq q < 1/2$. We develop parallel results for these
cases. In addition to a setting with noisy measurement outcomes,
these results can be used in a so called \emph{dilution model} (as observed
in \cite{atia09,CHKV:09}). In this
model, each item can be \emph{diluted} in each test with some a priori known
probability. In a network setting, this would correspond to the case where a test on a path with a
congested link can turn out to be negative with some probability. We
show that similar scaling results holds for this case as well.

\noindent \textbf{{\em Other group testing problems on graphs: }} Several
variations of classical group testing have been studied in the
literature that possess a graph theoretic nature. A notable example is
the problem of learning hidden sparse subgraphs (or more generally,
hypergraphs), defined as follows (cf.\ \cite{ref:hidden}): Assume
that, for a given graph, a small number of the edges are marked as
defective. The problem is to use a small number of measurements of the
following type to identify the set of defective edges: Each
measurement specifies a subset of vertices, and the outcome would be
positive iff the graph induced on the subset contains a defective edge. Another
variation concerns group testing with constraints defined by a rooted
tree. Namely, the set of items corresponds to the leaves of a given
rooted tree, and each test is restricted to pool all the leaves that
descend from a specified node in the tree (see
\cite[Chapter~12]{du}). To the best of our knowledge, our work is the
first variation to consider the natural restriction of the pools with
respect to the \emph{paths} on a given graph.

The rest of this paper is organized as follows. In
Section~\ref{sec:notation}, we introduce our notation and mention some
basic facts related to group testing and random walks on
graphs. Section~\ref{sec:results} formally describes the problem that
we consider and states our main results. In Section~\ref{sec:proof} we
prove the main results, and finally, in Section~\ref{sec:instan} show
instantiations of the result to the important cases of
graph-constrained group testing on regular expander graphs and random
graphs in the Erd\H{o}s-R\'enyi model.




\section{Definitions and Notation} \label{sec:notation} In this
section we introduce some tools, definition and notations which are
used throughout the paper.

\begin{defn}
  For two given boolean vectors $S$ and $T$ of the same length we
  denote their element-wise logical $\mathsf{or}$ by $S\vee T$. More
  generally, we will use $\bigvee_{i=1}^d S_i$ to denote the
  element-wise $\mathsf{or}$ of $d$ boolean vectors $S_1, \ldots, S_d$. The logical
  subtraction of two boolean vectors $S=(s_1, \ldots, s_n)$ and
  $T=(t_1, \ldots, t_n)$, denoted by $S\setminus T$, is defined as a
  boolean vector which has a $1$ at position $i$ if and only if
  $s_i=1$ and $t_i=0$. We also use $|S|$ to show the number of $1$'s
  in (i.e., the Hamming weight of) a vector $S$.
\end{defn}

We often find it convenient to think of boolean vectors as
characteristic vectors of sets. That is, $x \in \{0,1\}^n$ would
correspond to a set $X \subseteq [n]$ (where $[n] := \{1,\ldots, n\}$)
such that $i \in X$ iff the entry at the $i$th position of $x$ is
$1$. In this sense, the above definition extends the set-theoretic
notions of union, subtraction, and cardinality to boolean vectors.

Matrices that are suitable for the purpose of group testing are known
as \emph{disjunct} matrices. The formal definition is as follows.

\begin{defn}
  An $m \times n$ boolean matrix $M$ is called $d$-disjunct, if, for
  every column $S_0$ and every choice of $d$ columns $S_1, \ldots,
  S_d$ of $M$ (different from $S_0$), there is at least one row at
  which the entry corresponding to $S_0$ is $1$ and those
  corresponding to $S_1, \ldots, S_d$ are all zeros. More generally,
  for an integer $e \geq 0$, the matrix is called $(d, e)$-disjunct if
  for every choice of the columns $S_i$ as above, they satisfy
  \begin{align*}
    |S_0 \setminus \bigvee_{i=1}^d S_i| > e.
  \end{align*}
  A $(d,0)$-disjunct matrix is said to be $d$-disjunct.
\end{defn}

A classical observation in group testing theory states that disjunct
matrices can be used in non-adaptive group testing schemes to
distinguish sparse boolean vectors (cf.~\cite{du}).  More precisely,
suppose that a $d$-disjunct matrix $M$ with $n$ columns is used as the
measurement matrix; i.e., we assume that the rows of $M$ are the
characteristic vectors of the pools defined by the scheme. Then, the
test outcomes obtained by applying the scheme on two distinct
$d$-sparse vectors of length $n$ must differ in at least one
position. More generally, if $M$ is taken to be $(d,e)$-disjunct, the
test outcomes must differ in at least $e+1$ positions.  Thus, the more
general notion of $(d, e)$-disjunct matrices is useful for various
``noisy'' settings, where we are allowed to have a few false outcomes
(in particular, up to $\lfloor (e-1)/2 \rfloor$ incorrect measurement
outcomes can be tolerated by $(d, e)$-disjunct matrices without
causing any confusion).

For our application, sparse vectors (that are to be distinguished)
correspond to boolean vectors
encoding the set of defective vertices (or edges) in a given undirected
graph. 
The encoding is such that the coordinate positions are indexed by the
set of vertices (edges) of the graph and a position contains $1$
iff it corresponds to a defective vertex (edge).
Moreover, we aim to construct disjunct matrices that are also
constrained to be \emph{consistent} with the underlying graph.

\begin{defn}
  Let $G=(V, E)$ be an undirected graph, and $A$ and $B$ be boolean
  matrices with $|V|$ and $|E|$ columns, respectively.  The columns of
  $A$ are indexed by the elements of $V$ and the columns of $B$ are
  indexed by the elements of $E$. Then,
  \begin{itemize}
  \item The matrix $A$ is said to be \emph{vertex-consistent} with $G$
    if each row of $A$, seen as the characteristic vector of a subset
    of $V$, exactly represents the set of vertices visited by some
    walk on $G$.

  \item The matrix $B$ is said to be \emph{edge-consistent} with $G$
    if each row of $B$, seen as the characteristic vector of a subset
    of $E$, exactly corresponds to the set of edges traversed by a
    walk on $G$.
  \end{itemize}

  Note that the choice of the walk corresponding to each row of $A$ or
  $B$ need not be unique. Moreover, a walk may visit a vertex (or
  edge) more than once.
\end{defn}

\begin{defn}
  An undirected graph $G=(V,E)$ is called $(D, c)$-uniform, for some
  $c \geq 1$, if the degree of each vertex $v\in V$ (denoted by
  $\deg(v)$) is between $D$ and $cD$.
\end{defn}



\begin{defn}
  The \emph{point-wise distance} of two probability distributions
  $\mu, \mu'$ on a finite space $\Omega$ is defined as
  \[
  \| \mu - \mu' \|_\infty := \max_{i \in \Omega} |\mu(i) - \mu'(i)|,
  \]
  where $\mu(i)$ (resp., $\mu'(i)$) denotes the probability assigned by
  $\mu$ (resp., $\mu'$) to the outcome $i \in \Omega$.  We say that
  the two distributions are $\delta$-close if their point-wise
  distance is at most $\delta$.
\end{defn}

For notions such as random walks,
stationary distribution and mixing time we refer to
many text books on probability theory, Markov chains, and
randomized algorithms. In particular for an accessible treatment
of the basic notions, see \cite[Chapter~6]{ref:MR} or \cite[Chapter~7]{ref:MU}.
The particular variation of the mixing time that we will use in this
work is defined with respect to the point-wise distance as follows.

\begin{defn} \label{def:mixing} Let $G=(V,E)$ with $|V|=n$ be a $(D,
  c)$-uniform graph and denote by $\mu$ its stationary distribution.
  For $v\in V$ and an integer $\tau$, denote by $\mu^\tau_v$
  the distribution that a random walk of length $\tau$ starting at $v$
  ends up at. Then, the \emph{$\delta$-mixing time} of $G$ (with respect to the
  $\ell_\infty$ norm\footnote{Note that the mixing time highly depends on the
  underlying distance by which the distance between two distributions
  is quantified. In particular, we are slightly deviating from the 
  more standard definition which is with respect to the variation $(\ell_1)$ distance (see,
  e.g., \cite[Definition~11.2]{ref:MU}).}) is the smallest integer $t$ such that
  $\| \mu^\tau_v - \mu \|_\infty \leq \delta$, for $\forall \tau \geq t$ and
  $\forall v\in V$. For
  concreteness, we define the quantity $T(n)$ as the $\delta$-mixing
  time of $G$ for $\delta := (1/2cn)^2$.
\end{defn}
Throughout this work, the constraint graphs are considered to be $(D,
c)$-uniform, for an appropriate choice of $D$ and some (typically
constant) parameter $c$. When $c=1$, the graph is $D$-regular.

For a graph to have a small mixing time, a random walk starting from
any vertex must quickly induce a uniform distribution on the vertex
set of the graph. Intuitively this happens if the graph has no
``bottle necks'' at which the walk can be ``trapped'', or in other
words, if the graph is ``highly connected''.  The standard notion of
\emph{conductance}, as defined below, quantifies the connectivity of a
graph.

\begin{defn} \label{def:conductance} Let $G=(V,E)$ be a graph on $n$
  vertices. For every $S \subseteq V$, define $\Delta(S) := \sum_{v
    \in S} \deg(v)$, $\bar{S} := V \setminus S$, and denote by $E(S,
  \bar{S})$ the number of edges crossing the cut defined by $S$ and
  its complement. Then the \emph{conductance} of $G$ is defined by the
  quantity
  \[
  \Phi(G) := \min_{S \subseteq V\colon \Delta(S) \leq |E|} \frac{E(S,
    \bar{S})}{\Delta(S)}.
  \]
\end{defn}

We also formally define two important classes of graphs, for which
we will specialize our results.

\begin{defn}
\label{def:ERgraph}
Take a complete graph on $n$ vertices, and remove edges independently with probability $1 - p$. The resulting graph is called
the Erd\H{o}s-R\'enyi random graph, and denoted by $G(n,p)$.
\end{defn}

\begin{defn}
\label{def:EXPgraph}
For a graph $G=(V,E)$ with $|V|=n$, the (edge) expansion of $G$ is defined as
\[
h(G)=\min_{S \subseteq V\colon 0< |S| \leq \frac{n}{2}} \frac{E(S,\bar{S})}{|S|}.
\]
A family  $\mathcal{G}$ of $D$-regular graphs is called an (edge) expander family if there exists a constant $\sigma > 0$ such that  $h(G)\geq \sigma$ for each $G\in\mathcal{G}$. In particular each $G\in\mathcal{G}$ is called an expander graph.
\end{defn}

For a general study of Erd\H{o}s-R\'enyi random graphs and their
properties we refer to the fascinating book of Bollob\'as \cite{ref:BB}.
For the terminology on expander graphs, we refer the reader to the
excellent survey by Hoory, Linial and Wigderson \cite{HLW06}.

\begin{defn} \label{def:pi} Consider a particular random walk $W :=
  (v_0, v_1, \ldots, v_t)$ of length $t$ on a graph $G=(V,E)$, where
  the random variables $v_i \in V$ denote the vertices visited by the
  walk, and form a Markov chain.  We distinguish the following
  quantities related to the walk $W$:

  \begin{itemize}
  \item For a vertex $v \in V$ (resp., edge $e \in E$), denote by
    $\pi_v$ (resp., $\pi_e$) the probability that $W$ passes $v$
    (resp., $e$).

  \item For a vertex $v \in V$ (resp., edge $e \in E$) and subset $A
    \subseteq V$, $v \notin A$ (resp., $B \subseteq E$, $e \notin B$),
    denote by $\pi_{v,A}$ (resp., $\pi_{e,B}$) the probability that
    $W$ passes $v$ but none of the vertices in $A$ (resp., passes $e$
    but none of the edges in $B$).
  \end{itemize}

  Note that these quantities are determined by not only $v, e, A, B$
  (indicated as subscripts) but they also depend on the choice of the
  underlying graph, the distribution of the initial vertex $v_0$ and
  length of the walk $t$. However, we find it convenient to keep the
  latter parameters implicit when their choice is clear from the
  context.
\end{defn}

In the previous definition, the length of the random walk was taken as
a fixed parameter $t$.  Another type of random walks that we consider
in this work have their end points as a parameter and do not have an a
priori fixed length. In the following, we define similar probabilities
related to the latter type of random walks.

\begin{defn} \label{def:piu} Consider a particular random walk $W :=
  (v_0, v_1, \ldots, u)$ on a graph $G=(V,E)$ that continues until it
  reaches a fixed vertex $u \in V$. We distinguish the following
  quantities related to $W$:
  For a vertex $v \in V$ (resp., edge $e \in E$) and subset $A
  \subseteq V$, $v \notin A$ (resp., $B \subseteq E$, $e \notin B$),
  denote by $\piu_{v,A}$ (resp., $\piu_{e,B}$) the probability that
  $W$ passes $v$ but none of the vertices in $A$ (resp., passes $e$
  but none of the edges in $B$).

  Again these quantities depend on the choice of $G$ and the
  distribution of $v_0$ that we will keep implicit.
\end{defn}


\section{Problem setting and Main Results} \label{sec:results}

\textbf{Problem Statement. } Consider a given graph $G=(V,E)$ in which
at most $d$ vertices (resp., edges) are defective. The goal is to
characterize the set of defective items using a number of
measurements that is as small as possible, where each measurement determines whether
the set of vertices (resp., edges) observed along a path on the graph
has a non-empty intersection with the defective set. We call the
problem of finding defective vertices \emph{vertex group testing} and
that of finding defective edges \emph{edge group testing}.

\begin{figure}[t]
   \centering
   \includegraphics[width=8cm]{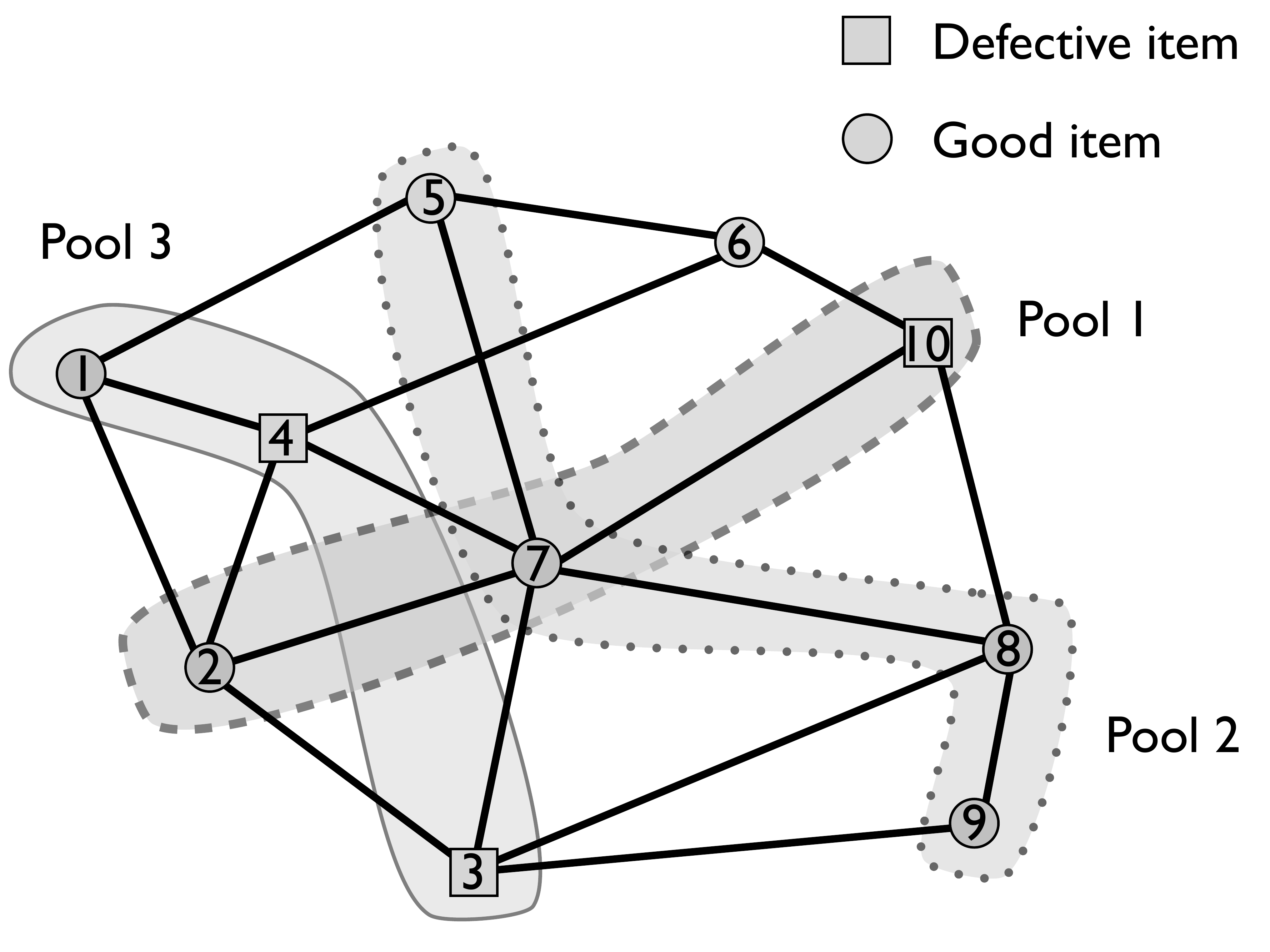}
   \caption{The result of pool~1 is positive since it contains a
     defective item, whereas the result of pool 2 is negative since it
     does not contain a defective item. Pool~3 is not consistent with
     the graph and thus not allowed since the items are not connected
     by a path.}\label{fig:graph}
 \end{figure}

As mentioned earlier, not all sets of vertices can be grouped
together, and only those that share a path on the underlying graph $G$
can participate in a pool (see Fig.~\ref{fig:graph}).

 In the following, we introduce four random constructions (designs)
 for both problems.  The proposed designs follow the natural idea of
 determining pools by taking random walks on the graph.

 \vspace{5mm}
 \noindent \framebox{
   \begin{minipage}[l]{\columnwidth-6mm}

     {\bf Design 1. }

     \begin{description}
     \item \emph{Given:} a constraint graph $G=(V,E)$ with $r \geq 0$
       designated vertices $s_1, \ldots, s_r \in V$, and integer
       parameters $m$ and $t$.
     \item \emph{Output:} an $m \times |V|$ boolean matrix $M$.
     \item \emph{Construction:} Construct each row of $M$
       independently as follows: Let $v \in V$ be any of the
       designated vertices $s_i$, or otherwise a vertex chosen
       uniformly at random from $V$.  Perform a random walk of length
       $t$ starting from $v$, and let the corresponding row of $M$ be
       the characteristic vector of the set of vertices visited by the
       walk.
     \end{description}
   \end{minipage}
 }

 \vspace{5mm}
 \noindent \framebox{
   \begin{minipage}[l]{\columnwidth-6mm}

     {\bf Design 2. }

     \begin{description}
     \item \emph{Given:} a constraint graph $G=(V,E)$ and integer
       parameters $m$ and $t$.
     \item \emph{Output:} an $m \times |E|$ boolean matrix $M$.
     \item \emph{Construction:} Construct each row of $M$
       independently as follows: Let $v \in V$ be any arbitrary vertex
       of $G$.  Perform a random walk of length $t$ starting from $v$,
       and let the corresponding row of $M$ be the characteristic
       vector of the set of edges visited by the walk.
     \end{description}
   \end{minipage}
 }

 \vspace{5mm}
 \noindent \framebox{
   \begin{minipage}[l]{\columnwidth-6mm}

     {\bf Design 3. }

     \begin{description}
     \item \emph{Given:} a constraint graph $G=(V,E)$ with $r \geq 0$
       designated vertices $s_1, \ldots, s_r \in V$, a \emph{sink
         node} $u \in V$, and integer parameter $m$.
     \item \emph{Output:} an $m \times |V|$ boolean matrix $M$.
     \item \emph{Constructions:} Construct each row of $M$
       independently as follows:
       Let $v \in V$ be any of the designated vertices $s_i$, or
       otherwise a vertex chosen uniformly at random from $V$.
       Perform a random walk starting from $v$ until we reach $u$, and
       let the corresponding row of $M$ be the characteristic vector
       of the set of vertices visited by the walk.
     \end{description}
   \end{minipage}
 }

 \vspace{5mm} \vspace{5mm}
 \noindent \framebox{
   \begin{minipage}[l]{\columnwidth-6mm}

     {\bf Design 4. }

     \begin{description}
     \item \emph{Given:} a constraint graph $G=(V,E)$, a \emph{sink
         node} $u \in V$, and integer parameter $m$.
     \item \emph{Output:} an $m \times |E|$ boolean matrix $M$.
     \item \emph{Construction:} Construct each row of $M$
       independently as follows: Let $v \in V$ be any arbitrary vertex
       of $G$.  Perform a random walk, starting from $v$ until we
       reach $u$, and let the corresponding row of $M$ be the
       characteristic vector of the set of edges visited by the walk.
     \end{description}
   \end{minipage}
 }\vspace{5mm}

 By construction, Designs 1~and~3 (resp., Designs 2~and~4) output
 boolean matrices that are vertex- (resp., edge-) consistent with the
 graph $G$. Our main goal is to show that, when the number of rows $m$
 is sufficiently large, the output matrices become $d$-disjunct (for a
 given parameter $d$) with overwhelming probability.

\begin{rem}
  Designs 1~and~3 in particular provide two choices for constructing
  the measurement matrix $M$. Namely, the start vertices can be chosen
  within a fixed set of designated vertices, or, chosen randomly among
  all vertices of the graph. As we will see later, in theory there is
  no significant difference between the two schemes. However, for some
  applications it might be the case that only a small subset of
  vertices are accessible as the starting points (e.g., in network
  tomography such a subset can be determined by the vantage points),
  and this can be modeled by an appropriate choice of the designated
  vertices in Designs 1~and~3.
\end{rem}

\begin{table}
  \caption{The asymptotic values of various parameters in Theorem~\ref{thm:disjunct}.}\label{parameter}
  \begin{center}
    \begin{tabular}{ |c | c | }
      \hline
      Parameter & Value\\\hline \hline
      $D_0$&$O(c^2 d T^2(n))$\\\hline
      $m_1, m_2$ & $O(c^4 d^2 T^2(n) \log(n/d))$ \\\hline
      $m_3$& $O(c^8 d^3 T^4(n) \log(n/d))$ \\\hline
      $m_4$& $O(c^9 d^3 D T^4(n) \log(n/d))$\\\hline
      $t_1$&$O(n/(c^3 dT(n)))$ \\\hline
      $t_2$&$O(n D/(c^3 dT(n)))$ \\\hline
      $e_1, e_2, e_3, e_4$&$\Omega(\eta d\log(n/d)/(1-\eta)^2)$\\\hline
      $m'_i, i \in [4]$&$O(m_i/(1-\eta)^2)$\\\hline
    \end{tabular}
  \end{center}
\end{table}

The following theorem states the main result of this
work, 
showing that our proposed designs indeed produce disjunct matrices
that can be used for the purpose of graph-constrained group testing.
We will state both noiseless results (corresponding to $d$-disjunct matrices),
and noisy ones (corresponding to $(d, e)$-disjunct ones, where the noise tolerance
$e$ depends on a fixed ``noise parameter'' $\eta \in [0,1)$). The proof of the
following theorem is given in Section~\ref{sec:proof}.

\begin{thm} \label{thm:disjunct} Let $\eta \geq 0$ be a fixed parameter,
  and suppose that $G=(V,E)$ is a $(D,c)$-uniform graph on $n$
  vertices with $\delta$-mixing time $T(n)$ (where $\delta :=
  (1/2cn)^2$).  Then there exist parameters with asymptotic values
  given in Table~\ref{parameter} such that, provided that $D \geq
  D_0$,

  \begin{enumerate}
  \item Design~1 with the path length $t := t_1$ and the number of
    measurements $m := m_1$ outputs a matrix $M$ that is
    vertex-consistent with $G$. Moreover, once the columns of $M$
    corresponding to the designated vertices $s_1, \ldots, s_r$ are
    removed, the matrix becomes $d$-disjunct with probability
    $1-o(1)$. More generally, for $m:=m'_1$ the matrix becomes $(d,
    e_1)$-disjunct with probability $1-o(1)$.
  \item Design~2 with path length $t := t_2$ and $m := m_2$
    measurements outputs a matrix $M$ that is edge-consistent with $G$
    and is $d$-disjunct with probability $1-o(1)$.  More generally,
    for $m:=m'_2$ the matrix becomes $(d, e_2)$-disjunct with
    probability $1-o(1)$.
  \item Design~3 with the number of measurements $m := m_3$ outputs a
    matrix $M$ that is vertex-consistent with $G$. Moreover, once the
    columns of $M$ corresponding to the designated vertices $s_1,
    \ldots, s_r$ and the sink node $u$ are removed, the matrix becomes
    $d$-disjunct with probability $1-o(1)$.  More generally, for
    $m:=m'_3$ the matrix becomes $(d, e_3)$-disjunct with probability
    $1-o(1)$.

  \item Design~4 with the number of measurements $m:=m_4$ outputs a
    matrix $M$ that is edge-consistent with $G$ and is $d$-disjunct
    with probability $1-o(1)$.  More generally, for $m:=m'_4$ the
    matrix becomes $(d, e_4)$-disjunct with probability $1-o(1)$.
  \end{enumerate}
\end{thm}

\begin{rem}
  In Designs 1~and~3, we need to assume that the designated vertices (if any)
  are not defective, and hence, their corresponding columns can be
  removed from the matrix $M$. By doing so, we will be able to ensure
  that the resulting matrix is disjunct. Obviously, such a restriction
  cannot be avoided since, for example, $M$ might be forced to contain
  an all-ones column corresponding to one of the designated vertices
  and thus, fail to be even $1$-disjunct.
\end{rem}

\begin{rem}
  By applying Theorem~\ref{thm:disjunct} on the complete graph (using
  Design~1), we get $O(d^2 \log (n/d))$ measurements, since in this
  case, the mixing time is $T(n)=1$ and also $c=1$. Thereby, we recover the trade-off
  obtained by the probabilistic construction in classical group
  testing (note that classical group testing corresponds to
  graph-constrained group testing on the vertices of the complete
  graph).
\end{rem}

We will show in Section~\ref{sec:instan} that, for our specific choice
of $\delta := (1/2cn)^2$, the $\delta$-mixing time of an
Erd\H{o}s-R\'enyi random graph $G(n,p)$ is (with
overwhelming
probability) $T(n) = O(\log n)$.  This bound more generally holds for
any graph with conductance $\Omega(1)$, and in particular, expander
graphs with constant spectral gap.  Thus we have the following result
(with a summary of the achieved parameters given in Table~\ref{tab:instan}).

\begin{thm}\label{thm:num-mes}
  There is an integer $D_0 = \Omega(d \log^2 n)$ such that for every
  $D \geq D_0$ the following holds: Suppose that the graph $G$ is
  either
  \begin{enumerate}
  \item A $D$-regular expander graph with normalized second largest
    eigenvalue (in absolute value) $\lambda$ that is bounded away from
    $1$; i.e., $\lambda = 1-\Omega(1)$, or,

  \item An Erd\H{o}s-R\'enyi random graph $G(n, D/n)$.
  \end{enumerate}
  Then for every $\eta \in [0,1)$, with probability $1-o(1)$ Designs
  1,~2,~3, and 4 output $(d,e)$-disjunct matrices (not considering the
  columns corresponding to the designated vertices and the sink in
  Designs~1 and~3), for some $e = \Omega(\eta d \log n)$, using
  respectively $m_1, m_2, m_3, m_4$ measurements, where $m_1,
  m_2=O(d^2 (\log^3 n)/(1-\eta)^2)$, $m_3=O(d^3 (\log^5 n)/(1-\eta)^2)$, and $m_4=O(d^3 D (\log^5 n)/(1-\eta)^2)$.
\end{thm}

\begin{table}
  \caption{The asymptotic values of the bounds achieved by Theorem~\ref{thm:num-mes}.}\label{tab:instan}
  \begin{center}
    \begin{tabular}{ |c | c | }
      \hline
      Parameter & Value\\\hline \hline
      $D_0$&$O(d \log^2 n)$\\\hline
      $m_1, m_2$ & $O(d^2 (\log^3 n)/(1-p)^2)$ \\\hline
      $m_3$& $O(d^3 (\log^5 n)/(1-p)^2)$ \\\hline
      $m_4$& $O(d^3 D (\log^5 n)/(1-p)^2)$\\\hline
      $e$ & $\Omega(\eta d \log n)$\\\hline
    \end{tabular}
  \end{center}
\end{table}

\noindent {\bf The fixed-input case. }
Recall that, as Theorem~\ref{thm:disjunct} shows, our proposed designs
almost surely produce disjunct matrices using a number of measurements
summarized in Table~\ref{parameter}. Thus, with overwhelming probability,
once we fix the resulting matrix, it has the combinatorial property of
distinguishing between \emph{any} two $d$-sparse boolean vectors (each corresponding
to a set of up to $d$ defective vertices, not including designated ones, for Designs 1~and~3, or
up to $d$ defective edges for Designs 2~and~4) in the \emph{worst case}.
However, the randomized nature of our designs can be used to our benefit to
show that, practically, one can get similar results with a number of measurements
that is almost by a factor $d$ smaller than what required by Theorem~\ref{thm:disjunct}.
Of course, assuming a substantially lower number of measurements, we should not
expect to obtain disjunct matrices, or equivalently, to be able to distinguish
between any two sparse vectors in the worst case. However, it can be shown
that, for every \emph{fixed} $d$-sparse vector $x$, the resulting matrix with
overwhelming probability will be able to distinguish between $x$ and any other $d$-sparse vector
using a lower number of measurements. In particular, with overwhelming probability
(over the choice of the measurements), from the measurement outcomes obtained
from $x$, it will be possible to uniquely reconstruct $x$. More precisely,
it is possible to show the following theorem, as proved in Section~\ref{sec:proof}.

\begin{thm} \label{thm:randomDisjunct}
Consider the assumptions of Theorem~\ref{thm:disjunct}, and let
$\gamma := (\log n)/(d \log(n/d))$. Consider any fixed set of
up to $d$ vertices $S \subseteq V$ such that $|S| \leq d$ and
$S \cap \{s_1, \ldots, s_r\} = \emptyset$ and any fixed set of
up to $d$ edges $T \subseteq E$, $|T| \leq d$. Then with probability
$1-o(1)$ over the randomness of the designs the following holds.

Let $M_1, \ldots M_4$ respectively denote the measurement matrices produced
by Designs $1, \ldots, 4$ with the number of rows set to $O(\gamma m'_i)$.
Then for every $S' \subseteq V$ and every $T' \subseteq E$ such that
$S' \neq S$, $T' \neq T$ and $|S'| \leq d$, $S' \cap \{s_1, \ldots, s_r\} = \emptyset$,
$|T'| \leq d$, we have that
\begin{enumerate}
 \item The measurement outcomes of $M_1$ on $S$ and $S'$ (resp.,
 $M_3$ on $S$ and $S'$) differ at more than $\Omega(\gamma e_1)$ (resp.,
 $\Omega(\gamma e_3)$) positions.

 \item The measurement outcomes of $M_2$ on $T$ and $T'$ (resp.,
 $M_4$ on $T$ and $T'$) differ at more than $\Omega(\gamma e_2)$ (resp.,
 $\Omega(\gamma e_4)$) positions.
\end{enumerate}
\end{thm}

A direct implication of this result is that (with overwhelming probability),
once we fix the matrices obtained
from our randomized designs with the lowered number of measurements (namely,
having $O(\gamma m'_i) \approx O(m'_i / d)$ rows), the fixed matrices will be able to distinguish
between \emph{almost all} pairs of $d$-sparse vectors (and in particular, uniquely
identify randomly drawn $d$-sparse vectors, with probability $1-o(1)$ over their
distribution).

\noindent {\bf Example in Network Tomography. }
Here we illustrate a simple concrete example that demonstrates how our
constructions can be used for network tomography in a simplified
model.  Suppose that a network (with known topology) is modeled by a
graph with nodes representing routers and edges representing links
that connect them, and it is suspected that at most $d$ links in the
network are congested (and thus, packets routed through them are
dropped).  Assume that, at a particular ``source node'' $s$, we wish
to identify the set of congested links by distributing packets that
originate from $s$ in the network.

First, $s$ generates a packet containing a time stamp $t$ and sends it
to a randomly chosen neighbor, who in turn, decrements the time stamp
and forwards the packet to a randomly chosen neighbor, etc.  The
process continues until the time stamp reaches zero, at which point
the packet is sent back to $s$ along the same path it has traversed.
This can be achieved by storing the route to be followed (which is
randomly chosen at $s$) in the packet. Alternatively, for practical
purposes, instead of storing the whole route in the packet, $s$ can
generate and store a random seed for a pseudorandom generator as a
header in the packet. Then each intermediate router can use the
specified seed to determine one of its neighbors to which the packet
has to be forwarded.

Using the procedure sketched above, the source node generates a number
of independent packets, which are distributed in the network. Each
packet is either returned back to $s$ in a timely manner, or,
eventually do not reach $s$ due to the presence of a congested link
within the route.  By choosing an appropriate timeout, $s$ can
determine the packets that are routed through the congested links.

The particular scheme sketched above implements our Design~2, and thus
Theorem~\ref{thm:disjunct} implies that, by choosing the number of
hops $t$ appropriately, after generating a sufficient number of
packets (that can be substantially smaller than the size of the
network), $s$ can determine the exact set of congested links.  This
result holds even if a number of the measurements produce false
outcomes (e.g., a congested link may nevertheless manage to forward a
packet, or a packet may be dropped for reasons other than congestion),
in which case by estimating an appropriate value for the \emph{noise
  parameter} $p$ in Theorem~\ref{thm:disjunct} and increasing the
number of measurements accordingly, the source can still correctly
distinguish the congested links. Of course one can consider different
schemes for routing the test packets. For example, it may be more
desirable to forward the packets until they reach a pre-determined
``sink node'', an approach that is modeled by our Designs 3~and~4
above.


\section{Proof of Theorems \ref{thm:disjunct}~and~\ref{thm:randomDisjunct}} \label{sec:proof}

Before discussing Theorem~\ref{thm:disjunct} and its proof, we
introduce some basic propositions that are later used in the
proof. The omitted proofs will be presented in the
appendix. Throughout this section, we consider an underlying graph
$G=(V,E)$ that is $(D, c)$-uniform, with mixing time $T(n)$ as in
Definition~\ref{def:mixing}.

\begin{prop} \label{lem:prob} Let $A, B_1, B_2, \ldots, B_n$ be events
  on a finite probability space, define $B := \cup_{i=1}^n B_i$, and
  suppose that:
  \begin{enumerate}
  \item For every $i \in [n]$, $\Pr[A \mid B_i] \leq \eps$.

  \item For every set $S \subseteq [n]$ with $|S| > k$, $\cap_{i \in
      S} B_i = \emptyset$.
  \end{enumerate}
  Then, $\Pr[A \mid B] \leq \eps k$.
\end{prop}
The proof of this proposition may be found in Section~\ref{pr:lem:prob}.
The following proposition is a corollary of a well-known result for the stationary distribution of irregular graphs \cite[Theorem~7.13]{ref:MU}.
A formal proof of this proposition is given in Section~\ref{pr:prop:stationary}.

\begin{prop} \label{prop:stationary}
  Let $G=(V,E)$ be a $(D, c)$-uniform graph, and
  denote by $\mu$ the stationary distribution of $G$ (assuming that $G$ is not bipartite). Then for each $v
  \in V$, $1/cn \leq \mu(v) \leq c/n$.
\end{prop}


\begin{prop} \label{lem:piV} For the quantities $\pi_v$ and $\pi_e$ in
  Definition~\ref{def:pi}, we have
  \[ \pi_v = \Omega\left(\frac{t}{cnT(n)}\right), \quad \pi_e =
  \Omega\left(\frac{t}{cDnT(n)}\right). \]
\end{prop}

The proof of this Proposition~\ref{lem:piV} is presented in
Section~\ref{pr:lem:piV}. In fact, a stronger statement than this proposition can be
obtained, that with noticeable probability, every fixed vertex (or
edge) is hit by the walk at least once but not too many times nor too
``early''. This is made more precise in the following two
propositions, which are proved in Sections~\ref{pr:prop:toomany} and
 \ref{pr:prop:early}, respectively.

\begin{prop} \label{prop:toomany} Consider any walk $W$ in Design~1
  (resp., Design~2).  There is a $k = O(c^2 T(n))$ such that, for
  every $v \in V$ and every $e \in E$, the probability that $W$ passes
  $v$ (resp., $e$) more than $k$ times is at most $ \pi_v/4$ (resp.,
  $\pi_e/4$).
\end{prop}

\begin{prop} \label{prop:early} For any random walk $W$ in Design~1, let $v
  \in V$ be any vertex that is not among the designated vertices $s_1,
  \ldots, s_r$. Then the probability that $W$ visits $v$ within the
  first $k$ steps is at most $k/D$.
\end{prop}

The following proposition shows that the distributions of two vertices
on a random walk that are far apart by a sufficiently large number of steps are almost
independent. The proof of this proposition may be found in Section~\ref{pr:prop:influence}.

\begin{prop} \label{prop:influence} Consider a random walk $w := (v_0,
  v_1, \ldots, v_t)$ on $G$ starting from an arbitrary vertex, and
  suppose that $j \geq i + T(n)$. Let $\cE$ denote any event that only
  depends on the first $i$ vertices visited by the walk. Then for every $u, v
  \in V$,
  \[
  | \Pr[v_i = u | v_j = v, \cE] - \Pr[v_i = u | \cE] | \leq 2/(3cn).
  \]
\end{prop}

The following lemmas, which form the technical core of this work,
lower bound the quantities $\pi_{v,A}$, $\pi_{e,B}$, $\piu_{v,A}$,
$\piu_{e,B}$ as defined by Definitions \ref{def:pi}~and~\ref{def:piu}.

\begin{lem} \label{lem:pi} There is a $D_0 = O(c^2 d T^2(n))$ and $t_1
  = O(n/(c^3 dT(n)))$ such that whenever $D \geq D_0$, by setting the
  path lengths $t := t_1$ in Design~1 the following holds.  Let $v \in
  V$, and $A \subseteq V$ be a set of at most $d$ vertices in $G$ such
  that $v \notin A$ and $A \cup \{v\}$ does not include any of the
  designated vertices $s_1, \ldots s_r$. Then
  \begin{equation} \label{eqn:pi} \pi_{v,A} = \Omega\left(\frac{1}{c^4
        d T^2(n)}\right).
  \end{equation}
\end{lem}

\begin{IEEEproof}
  Denote by $\mu$ the stationary distribution of $G$.  We know from
  Proposition~\ref{prop:stationary} that for each $u \in V$, $1/cn
  \leq \mu_u \leq c/n$.

  Let $k=O(c^2 T(n))$ be the quantity given by
  Proposition~\ref{prop:toomany}, $\cB$ denote the \emph{bad event}
  that $W$ hits some vertex in $A$. Moreover, let $\cG$ denote the
  \emph{good event} that $W$ hits $v$ no more than $k$ times in total and never
  within the first $2T(n)$ steps.  The probability of $\cG$ is, by
  Propositions \ref{prop:toomany}~and~\ref{prop:early}, at least
  \begin{displaymath}
\Pr(\cG)\geq 1-2T(n)/D-O(t/cnT(n)),
\end{displaymath}
   which can be made arbitrarily close to $1$
  (say larger than $0.99$) by choosing $D$ sufficiently large and $t$
  sufficiently small (as required by the statement). Now,
  \begin{align}
  \pi_{v,A} &= \Pr[\lnot \cB, v \in W] \nonumber\\
  &\geq \Pr[\lnot \cB, v \in W, \cG] \nonumber\\
  &= \Pr[v \in W, \cG] (1-\Pr[\cB \mid v \in W, \cG]).\label{eqn:initialbound}
  \end{align}

  By taking $D$ large enough, and in particular, $D = \Omega(c^2 d
  T^2(n))$, we can ensure that \[ 2T(n)/D \leq \pi_v/4.\] Combined with
  Proposition~\ref{prop:toomany}, we have $\Pr[v \in W, \cG] \geq
  \pi_v/2$, since
  \begin{eqnarray*}
    \Pr[v \in W, \cG] &=& \Pr[v \in W] + \Pr[\cG] - \Pr[(v \in W) \cup \cG] \\
    &\geq& \pi_v + (1-\pi_v/2) - 1 = \pi_v/2.
  \end{eqnarray*}
  Thus, \eqref{eqn:initialbound} gives
  \begin{equation} \label{eqn:secondbound} \pi_{v,A} \geq \pi_v
    (1-\Pr[\cB \mid v \in W, \cG])/2.
  \end{equation}
  Now we need to upperbound $\pi := \Pr[\cB \mid v \in W, \cG]$.
  Before doing so, fix some $i > 2T(n)$, and assume that $v_i = v$.
  Moreover, fix some vertex $u \notin A$ and assume that $v_0 = u$.
  We first try to upperbound $\Pr[\cB \mid v_i = v, v_0 = u]$.

  Let $\ell := i-T(n)$ and $\rho := i+T(n)$, and for the moment,
  assume that $T(n)+1 < \ell < \rho < t$ (a ``degenerate'' situation
  occurs when this is not the case). Partition $W$ into four parts:
  \begin{align*}
 W_1 &:= (v_0, v_1, \ldots, v_{T(n)}),\\
 W_2 &:= (v_{T(n)+1}, v_{T(n)+2}, \ldots, v_{\ell-1}),\\
 W_3 &:= (v_\ell, v_{\ell+1}, \ldots, v_\rho),\\
 W_4 &:= (v_{\rho+1}, v_{\rho+2}, \ldots, v_t).
\end{align*}
 For $j = 1,2,3,4$, define
 \begin{displaymath}
 \pi_j := \Pr[\text{$W_j$ enters $A$} \mid
  v_i=v, v_0=u].
  \end{displaymath}
  Now we upperbound each of the $\pi_j$. In a
  degenerate situation, some of the $W_i$ may be empty, and the
  corresponding $\pi_j$ will be zero.

  Each of the sub-walks $W_2$ and $W_4$ are ``oblivious'' of the
  conditioning on $v_i$ and $v_0$ (because they are sufficiently far
  from both and Proposition~\ref{prop:influence} applies).  In particular, the distribution of each vertex on $W_4$
  is point-wise close to $\mu$. Therefore, under our conditioning the
  probability that each such vertex belongs to $A$ is at most $|A|(c/n
  + \delta) < 2dc/n$. The argument on $W_2$ is similar, but more care
  is needed.  Without the conditioning on $v_i$, each vertex on $W_2$
  has an almost-stationary distribution. Moreover, by
  Proposition~\ref{prop:influence}, the conditioning on $v_2$ changes
  this distribution by up to $\delta' := 2/(3cn) < 1/n$ at each point.
  Altogether, for each $j \in \{T(n)+1, \ldots, \ell-1\}$, we have
  \begin{align*}
  \Pr[v_j \in A \mid v_i=v, v_0=u] &\leq |A|(c/n+\delta+\delta') \\
  &\leq  2dc/n.
\end{align*}
Using a union bound on the number of steps, we conclude that $\pi_2 + \pi_4 \leq 2dct/n$.

  In order to bound $\pi_3$, we observe that of all $D$ or more
  neighbors of $v_i$, at most $d$ can lie on $A$.  Therefore,
\begin{displaymath}
  \Pr[v_{i+1} \in A \mid v_i=v, v_0=u] \leq d/D.
\end{displaymath}
Similarly,
\begin{displaymath}
  \Pr[v_{i+2} \in A \mid v_i=v,v_0=u,v_{i+1}] \leq d/D,
\end{displaymath}
regardless
  of $v_{i+1}$ which means
\begin{displaymath}
\Pr[v_{i+2} \in A \mid v_i=v,v_0=u] \leq d/D,
\end{displaymath}
 and in general,
\begin{equation} \label{eqn:inGeneral:a}
(\forall j = i+1, \ldots, \rho),\quad \Pr[v_{j} \in A \mid v_i=v,v_0=u] \leq d/D.
\end{equation}

Similarly we have,
 \begin{displaymath}
\Pr[v_{i-1} \in A \mid v_i=v] \leq
  d/D,
\end{displaymath}
  and by Proposition~\ref{prop:influence} (and
  time-reversibility), conditioning on $v_0$ changes this probability
  by at most $d\delta'$. Therefore,
\begin{displaymath}
 \Pr[v_{i-1} \in A \mid
  v_i=v,v_0=u] \leq d/D + d\delta',
\end{displaymath}
 and in general,
\begin{multline} \label{eqn:inGeneral:b}
 (\forall j = \ell, \ldots, i-1), \\ \quad \Pr[v_{j} \in A \mid v_i=v,v_0=u] \leq d/D + d\delta'.
\end{multline}
Altogether, using a union bound and by combining \eqref{eqn:inGeneral:a} and \eqref{eqn:inGeneral:b}, we get that
\begin{displaymath}
\pi_3 \leq 2dT(n)/D + dT(n)/n \leq 3dT(n)/D.
  \end{displaymath}
  Using the same reasoning,
  $\pi_1$ can be bounded as
\begin{displaymath}
  \pi_1 \leq dT(n)/D + d T(n)/n \leq 2dT(n)/D.
\end{displaymath}
 Finally, we
  obtain
  \begin{align}
 \Pr[\cB \mid v_i = v, v_0 = u] &\leq \pi_1+\pi_2+\pi_3+\pi_4\nonumber \\
 &\leq
  \frac{5dT(n)}{D}+\frac{2dct}{n}.
\end{align}
Our next step is to relax the conditioning on the starting point of the walk.
  The probability that the initial vertex is in $A$ is at most $d/n$
  (as this happens only when the initial vertex is taken randomly),
  and by Proposition~\ref{prop:influence}, conditioning on $v_i$
  changes this probability by at most $d\delta' < d/n$. Now we write
  \begin{align*}
    \Pr[\cB \mid v_i = v] &\leq \Pr[v_0 \in A] + \Pr[\cB \mid v_i = v, v_0 \notin A] \\
    &\leq \Pr[v_0 \in A] + \pi_1+\pi_2+\pi_3+\pi_4 \\
    &\leq \frac{5dT(n)}{D}+\frac{4dct}{n},
  \end{align*}
where we have used the chain rule in the first inequality, and Proposition~\ref{lem:prob}
with $k=1$ for the second one.
  Now, since $\Pr[\cG]$ is very close to $1$, conditioning on this
  event does not increase probabilities by much (say no more than a
  factor $1.1$). Therefore,
  \[
  \Pr[\cB \mid v_i = v, \cG] \leq 1.1
  \left(\frac{5dT(n)}{D}+\frac{4dct}{n}\right).
  \]
  Now in the probability space conditioned on $\cG$, define events
  $\cG_i$, $i = 2T(n)+1, \ldots t$, where $\cG_i$ is the event that
  $v_i = v$. Note that the intersection of more than $k$ of the
  $\cG_i$ is empty (as conditioning on $\cG$ implies that the walk
  never passes $v$ more than $k$ times), and moreover, the union of
  these is the event that the walk passes $v$. Now we apply
  Proposition~\ref{lem:prob} to conclude that
  \begin{align*}
    \Pr[\cB \mid v \in W, \cG]
    &\leq 1.1 k \left(\frac{5dT(n)}{D}+\frac{4dct}{n}\right) \\
    &= O\left( c^2 T(n) \left(\frac{5dT(n)}{D}+\frac{4dct}{n}\right) \right).
  \end{align*}
  By taking $D=\Omega(c^2 d T^2(n))$ and $t=O(n/c^3 dT(n))$ we can
  make the right hand side arbitrarily small (say at most $1/2$).  Now
  we get back to \eqref{eqn:secondbound} to conclude, using
  Proposition \ref{lem:piV}, that
  \[
  \pi_{v,A} \geq \pi_v/4 = \Omega\left(\frac{t}{cnT(n)}\right) =
  \Omega\left(\frac{1}{c^4 d T^2(n)}\right).
  \]
\end{IEEEproof}

Similarly, we can bound the edge-related probability $\pi_{e,B}$ as in
the following lemma. The proof of the lemma is very similar to that of
Lemma~\ref{lem:pi}, and is therefore skipped for brevity.

\begin{lem} \label{lem:piE} There is a $D_0 = O(c^2 d T^2(n))$ and
  $t_2 = O(nD/c^3 dT(n))$ such that whenever $D \geq D_0$, by setting
  the path lengths $t := t_2$ in Design~2 the following holds.  Let $B
  \subseteq E$ be a set of at most $d$ edges in $G$, and $e \in E$, $e
  \notin B$. Then
  \begin{equation} \label{eqn:piE} \pi_{e,B} =
    \Omega\left(\frac{1}{c^4 d T^2(n)}\right).
  \end{equation}
\end{lem}

In Designs 3~and~4, the quantities $\piu_{v,A}$ and $\piu_{e,B}$
defined in Definition~\ref{def:piu} play a similar role as $\pi_{v,A}$
and $\pi_{e,B}$. In order to prove disjunctness of the matrices
obtained in Designs 3~and~4, we will need lower bounds on $\piu_{v,A}$
and $\piu_{e,B}$ as well. In the following we show the desired lower
bounds.

\begin{lem} \label{lem:piu} There is a $D_0 = O(c^2 d T^2(n))$ such
  that whenever $D \geq D_0$, in Design~3 the following holds.  Let $v
  \in V$, and $A \subseteq V$ be a set of at most $d$ vertices in $G$
  such that $v \notin A$ and $A \cup \{v\}$ is disjoint from $\{s_1,
  \ldots s_r, u\}$. Then
  \begin{equation} \label{eqn:piu} \piu_{v,A} =
    \Omega\left(\frac{1}{c^8 d^2 T^4(n)}\right).
  \end{equation}
\end{lem}

\begin{IEEEproof}
  Let $D_0$ and $t_1$ be quantities given by Lemma\footnote{In fact,
    as will be clear by the end of the proof, Lemma~\ref{lem:pi}
    should be applied with the sparsity parameter $d+1$ instead of
    $d$. However, this will only affect constant factors that we
    ignore.}~\ref{lem:pi}. Let $w_0$ denote the start vertex of a walk
  performed in Design~3, and consider an infinite walk $W=(v_0, v_1,
  v_2, \ldots)$ that starts from a vertex identically distributed with
  $w_0$.  Let the random variables $i, j, k$ respectively denote the
  times that $W$ visits $v, u$, and any of the vertices in $A$ for the
  first time. Therefore, $v_i = v$, $v_j = u$, and $v_k \in A$, $v_t
  \neq v$ for every $t < i$ and so on.  Then the quantity $\piu_{v,A}$
  that we wish to bound corresponds to the probability that $i < j <
  k$, that is, probability of the event that in $W$, the first visit
  of $v$ occurs before the walk reaches the sink node $u$ for the first time,
  and moreover, the walk never hits $A$ before reaching $u$.
  Observe that this event in particular contains the sub-event that $i
  \leq t_1$, $t_1 < j \leq 2t_1$, and $k > 2t_1$, where $t_1$ is picked
  as in Lemma~\ref{lem:pi}.  Denote by $\good
  \subseteq V^{t_1+1}$ the set of all sequences of $t_1+1$ vertices of
  $G$ (i.e., walks of length $t_1$) that include $v$ but not any of
  the vertices in $A \cup \{u\}$.  Now, we can write

\begin{eqnarray}
  \piu_{v,A} &=& \Pr[i < j < k] \\ &\geq& \Pr[i \leq t_1 < j \leq 2t_1 < k] \nonumber \\
  &=& \Pr[(i \leq t_1) \land (j > t_1) \land (k > t_1)] \cdot \nonumber \\ && \Pr[t_1 < j \leq 2t_1 < k \mid \nonumber \\
  && \qquad (i \leq t_1) \land (j > t_1) \land (k > t_1)] \nonumber \\
  &=& \Pr[(v_0, \ldots, v_{t_1}) \in \good] \cdot \nonumber \\ && \Pr[t_1 < j \leq 2t_1 < k \mid (v_0, \ldots, v_{t_1}) \in \good] \nonumber \\ \quad \label{eqn:pipi}
\end{eqnarray}


The probability $\Pr[(v_0, \ldots, v_{t_1}) \in \good]$ is exactly
$\pi_{v,{A \cup \{u\}}}$ with respect to the start vertex
$w_0$. Therefore, Lemma~\ref{lem:pi} gives the lower bound
\[ \Pr[(v_0, \ldots, v_{t_1}) \in \good] = \Omega\left(\frac{1}{c^4 d
    T^2(n)}\right).\] Furthermore observe that, regardless of the
outcome $(v_0, \ldots, v_{t_1}) \in \good$, we have
\begin{displaymath}
\Pr[t_1 <
j \leq 2t_1 < k \mid v_0, \ldots, v_{t_1}]=\pi_{u,A}
\end{displaymath}
where $\pi_{u,A}$ is taken with
respect to the start vertex $v_{t_1}$. Therefore, since $v_{t_1}
\notin A \cup \{u\}$, again we can use Lemma~\ref{lem:pi} to conclude
that
\[
\Pr[t_1 < j \leq 2t_1 < k \mid (v_0, \ldots, v_{t_1}) \in \good]=
\Omega\left(\frac{1}{c^4 d T^2(n)}\right).
\]
By plugging the bounds in \eqref{eqn:pipi} the claim follows.
\end{IEEEproof}

A similar result can be obtained for Design~4 on the edges. Since the
arguments are very similar, we only sketch a proof.

\begin{lem} \label{lem:piuE} There is a $D_0 = O(c^2 d T^2(n))$ such
  that whenever $D \geq D_0$, in Design~4 the following holds.  Let $B
  \subseteq E$ be a set of at most $d$ edges in $G$, and $e \in E$, $e
  \notin B$. Then
  \begin{equation} \label{eqn:piuE} \piu_{e,B} =
    \Omega\left(\frac{1}{c^9 d^2 D T^4(n)}\right).
  \end{equation}
\end{lem}

\begin{IEEEproof} {\em (sketch) }
  Similar to the proof of Lemma~\ref{lem:piu}, we consider an infinite
  continuation $W=(v_0, v_1, \ldots)$ of a walk performed in Design~4
  and focus on its first $t_1 + t_2$ steps, where $t_1$ and $t_2$ are
  respectively the time parameters given by Lemmas
  \ref{lem:pi}~and~\ref{lem:piE}. Let
  \begin{align*}
W_1 &:= (v_0, \ldots, v_{t_1}),\\
 W_2 &:= (v_{t_1+1}, \ldots, v_{t_1+t_2}).
\end{align*}
 Again following the
  argument of Lemma~\ref{lem:piu}, we lower bound $\piu_{e,B}$ by the
  probability of a sub-event consisting the intersection of the
  following two events:
  \begin{enumerate}
  \item The event $\cE_1$ that $W_1$ visits $e$ but neither the sink
    node $u$ nor any of the edges in $B$, and
  \item The event $\cE_2$ that $W_2$ visits the sink node $u$ but none
    of the edges in $B$.
  \end{enumerate}
  Consider the set $A \subseteq V$ consisting of the endpoints of the
  edges in $B$ and denote by $v \in V$ any of the endpoints of
  $e$. Let $p := \pi_{v,A}$ (with respect to the start vertex $v_0$).
  Now, $\Pr[\cE_1] \geq p/(cD)$ since upon visiting $v$, there is a
  $1/\deg(v)$ chance that the next edge taken by the walk turns out to
  be $e$. The quantity $p$ in turn, can be lower bounded using
  Lemma~\ref{lem:pi}.  Moreover, regardless of the outcome of $W_1$,
  the probability that $W_2$ visits $u$ but not $B$ (and subsequently,
  the conditional probability $\Pr[\cE_2 \mid \cE_1]$) is at least the
  probability $\pi_{e',B}$ (with respect to the start vertex
  $v_{t_1}$), where $e' \in E$ can be taken as any edge incident to
  the sink node $u$.  This latter quantity can be lower bounded using
  Lemma~\ref{lem:piE}. Altogether, we obtain the desired lower bound
  on $\piu_{e,B}$.
\end{IEEEproof}

\begin{rem} \label{rem:question} It is natural to ask whether the
  exponent of $d^2$ in the denominator of the lower bound in
  Lemma~\ref{lem:piu} can be improved. We argue that this is not the
  case in general, by considering the basic where the underlying graph is the
  complete graph $K_n$ and each walk is performed starting from a
  random node.
  Consider an infinite walk $W$ starting at a random vertex and moreover, the
  set of $d+2$ vertices $A' := A \cup \{u,v\}$.  Due to the symmetry
  of the complete graph, we expect that the order at which $W$ visits
  the vertices of $A'$ for the first time is uniformly distributed
  among the $(d+2)!$ possible orderings of the elements of $A'$.
  However, in the event corresponding to $\piu_{v,A}$, we are
  interested in seeing $v$ first, then $u$, and finally the elements
  of $A$ in some order.  Therefore, for the case of complete graph we know that
  $\piu_{v,A} = O(1/d^2)$, and thus, the quadratic dependence on $d$
  is necessary even for very simple examples.
\end{rem}
\begin{rem} Another question concerns the dependence of the lower
  bound in Lemma~\ref{lem:piuE} on the degree parameter $D$. Likewise
  Remark~\ref{rem:question}, an argument for the case of complete
  graph suggests that in general this dependence cannot be
  eliminated. For edge group testing on the complete graph, we expect
  to see a uniform distribution on the ordering at which we visit a
  particular set of edges in the graph. Now the set of edges of our
  interest consists of the union of the set $B \cup \{e\}$ and all the
  $n-1$ edges incident to the sink node $u$, and is thus of size
  $n+d$. The orderings that contribute to $\piu_{e,B}$ must have $e$
  as the first edge and an edge incident to $u$ as the second edge.
  Therefore we get that, for the case of complete graph,
 \begin{displaymath}
 \piu_{e,B} =
  O(1/n) = O(1/D),
\end{displaymath}
  which exhibits a dependence on the degree in the
  denominator.
\end{rem}
Now, we are ready to prove our main theorem.
\begin{IEEEproof}[Proof of Theorem~\ref{thm:disjunct}]
  We prove the first part of the theorem. Proofs of the other parts
  follow the same reasoning. The high-level argument is similar to the well known
  probabilistic argument in classical group testing, but we will have
  to use the tools that we have developed so far for working out the
  details. By construction, the output matrix $M$ is vertex-consistent
  with $G$.  Now, take a vertex $v \in V$ and $A \subseteq V$ such
  that $v \notin A$, $|A| \leq d$, and $(\{v\} \cup A) \cap \{s_1,
  \ldots, s_r\} = \emptyset$.  For each $i = 1, \ldots m_1$, define a
  random variable $X_i \in \{0,1\}$ such that $X_i = 1$ iff the $i$th
  row of $M$ has a $1$ entry at the column corresponding to $v$ and all-zeros
  at those corresponding to the elements of $A$.  Let $X :=
  \sum_{i=1}^{m_1} X_i$. Note that the columns corresponding to $v$
  and $A$ violate the disjunctness property of $M$ iff $X = 0$, and
  that the $X_i$ are independent Bernoulli random variables. Moreover,
  \begin{displaymath}
  \bbE[X_i] = \Pr[X_i = 1] = \pi_{v,A},
\end{displaymath}
 since $X_i = 1$ happens
  exactly when the $i$th random walk passes vertex $v$ but never hits
  any vertex in $A$. Now by using Lemma~\ref{lem:pi} we can ensure that,
  for an appropriate choice of $D_0$ and $t_1$ (as in the statement of the lemma), we have
  $\pi_{v,A} = \Omega(1/(c^4dT^2(n)))$.

  Denote by $p_f$ the \emph{failure probability}, namely that the
  resulting matrix $M$ is not $d$-disjunct.  By a union bound we get
  \begin{align*}
    p_f &\leq \sum_{v,A} (1-\pi_{v,A})^{m_1} \\
    &\leq \exp\left(d\log\frac{n}{d}\right) \cdot \left(1-\Omega\left(\frac{1}{c^4dT^2(n)}\right)\right)^{m_1}.
  \end{align*}
  Thus by choosing
  \begin{displaymath}
  m_1 = O\left(d^2 c^4 T^2(n)\log\frac{n}{d} \right)
\end{displaymath}
 we can ensure
  that $p_f = o(1)$, and hence, $M$ is $d$-disjunct with overwhelming
  probability.

  For the claim on $(d, e_1)$-disjunctness, note that a failure occurs
  if, for some choice of the columns (i.e., some choice of $v, A$), we have $X \leq e_1$.  Set
\begin{displaymath}
  \eta'
  := \eta \pi_{v,A} = \Omega\left(\frac{\eta}{c^4 d T^2(n)}\right),
\end{displaymath}
 and $e_1 := \eta'
  m'_1$. Note that $\bbE[X] = \pi_{v,A} m'_1$. Now by a Chernoff bound, we get
  \begin{align*}
  \Pr[X \leq \eta' m'_1] &\leq \exp\left( -\frac{(\bbE[X]-\eta' m'_1)^2}{2
      \bbE[X]} \right)\\
      & = \exp(-\bbE[X](1-\eta)^2/2).
\end{align*}

  So now, by a union bound, the failure probability $p_f$ becomes
  \begin{equation} \label{eqn:union}
  p_f \leq \exp\left(d\log\frac{n}{d}-m'_1(1-\eta)^2 \tilde{\pi}/2\right),
  \end{equation}
  where $\tilde{\pi}$ is the lower bound $\Omega(1/(c^4 d T^2(n)))$ on $\pi_{v,A}$.
  Thus we will have $p_f = o(1)$ by choosing
  \begin{displaymath}
  m'_1 = O\left(d^2 \log\frac{n}{d} c^4 T^2(n)
  / (1-\eta)^2\right).
\end{displaymath}
\end{IEEEproof}


\begin{IEEEproof}[Proof of Theorem~\ref{thm:randomDisjunct}]
The proof follows line-by-line the same arguments as in the proof of Theorem~\ref{thm:disjunct},
except that for the last union bound it would suffice to enumerate a substantially lower
number of choices of $v$ and $A$. In particular, consider Design~1 as in the proof of Theorem~\ref{thm:disjunct} (the argument for
the other designs is essentially the same). Then the only part of the proof that needs to be changed
is the union bound from which \eqref{eqn:union} follows.
Contrary to the proof of Theorem~\ref{thm:disjunct}, in the case we consider here, only up to $n$ choices
of the tuple $(v,A)$, in particular the following set, need to be enumerated:
\[
\cB := \{(v,A)\colon v \in V \setminus \{s_1, \ldots, s_r\}, A = S \setminus \{ v \} \}.
\]
Now assume that the resulting matrix ``satisfies'' all the choices of the tuples $(v, A) \in \cB$,
in that it has enough rows at which the entry corresponding to $v$ is $1$ while those
corresponding to $A$ are all zeros (this is guaranteed to hold, with
overwhelming probability, by the union bound).

Consider the case where $S' \nsubseteq S$ and take any $v \in S' \setminus S$. Since
$(v,S) \in \cB$, we can be sure that the measurement outcome corresponding to $S'$ would be
positive at more than $\Omega(\gamma e_1)$ of the positions while at those positions,
the outcome of $S$ must be zero. A similar argument is true for the case $S' \subseteq S$,
in which case it would suffice to take any $v \in S \setminus S'$ and observe that
$(v, S \setminus \{v\}) \in \cB$.

Altogether, from the above observations, the estimate \eqref{eqn:union} can be improved to
\[
  p_f \leq \exp\left(\log n-\tilde{m} (1-\eta)^2 \tilde{\pi}/2\right),
\]
where $\tilde{m}$ is the number of measurements. Therefore, we can ensure that $p_f = o(1)$ by
taking $\tilde{m} = O(\gamma m'_1)$, i.e., a factor $\gamma$ less than what needed by
Theorem~\ref{thm:disjunct}.
\end{IEEEproof}

\section{Proof of Theorem \ref{thm:num-mes}}\label{sec:instan}

In Theorem \ref{thm:num-mes} we consider two important instantiations
of the result given by Theorem~\ref{thm:disjunct}, namely when $G$ is
taken as an expander graph with constant spectral gap, and when it is
taken as an Erd\H{o}s-R\'enyi random graph $G(n,p)$. In the following
we show that in both cases (and provided that $p$ is not too small),
the mixing time is $O(\log n)$ (with probability $1-o(1)$).  Then Theorem~\ref{thm:disjunct} will
lead to the proof.

Before we proceed, we need to bound the distance between the
stationary distribution and the distribution obtained after $t$ random
steps on a graph. The following theorem, which is a direct corollary
of a result in \cite{ref:mixing}, is the main tool that we will need. We
skip the proof of this theorem here and refer to the main article for interested
readers.

\begin{thm}[\cite{ref:mixing}] \label{thm:JS} Let $G$ be an undirected graph with
  stationary distribution $\mu$, and denote by $d_\mathrm{min}$ and
  $d_\mathrm{max}$ the minimum and maximum degrees of its vertices,
  respectively. Let $\mu_v^t$ be the distribution obtained by any random
  walk on $G$ in $t$ steps starting at node $v$.  Then for all $v\in V$
  \begin{align*}
  \| \mu_v^t - \mu \|_\infty \leq (1-\Phi(G)^2/2)^t d_\mathrm{max} /
  d_\mathrm{min},
  \end{align*}
  where $\Phi(G)$ denotes the conductance of $G$ as in Definition~\ref{def:conductance}.
\end{thm}

\subsection{The Erd\H{o}s-R\'enyi random graph}

First, we present some tools for the case of random graphs.  Consider
a random graph $G(n,p)$ which is formed by removing each edge of the
complete graph on $n$ vertices independently with probability
$1-p$. Our focus will be on the case where $np \gg \ln n$. In this
case, the resulting graph becomes (almost surely) connected and the degrees are
highly concentrated around their expectations. In particular, we can
show the following fact, which we believe to be folklore. The proof of
this is presented in Section~\ref{pf:prop:randomDegree}.

\begin{prop} \label{prop:randomDegree} For every $\eps > 0$, with
  probability $1-o(1)$, the random graph $G(n,p)$ with $np \geq
  (2/\eps^2) \ln n $ is $(np(1-\eps), (1+\eps)/(1-\eps))$-uniform.
\end{prop}

In light of Theorem~\ref{thm:JS}, all we need to show is a lower bound on
the conductance of a random graph. This is done in the following.

\begin{lem} \label{lem:randomExp} For every $\varphi < 1/2$, there is
  an $\alpha > 0$ such that a random graph $G=G(n,p)$ with $p \geq
  \alpha \ln n/n$ has conductance $\Phi(G) \geq \varphi$ with
  probability $1-o(1)$.
\end{lem}
\begin{IEEEproof}
  First, note that by Proposition~\ref{prop:randomDegree} we can
  choose $\alpha$ large enough so that with probability $1-o(1)$, the
  degree of each vertex in $G$ is between $D(1-\eps)$ and $D(1+\eps)$,
  for an arbitrarily small $\eps > 0$ and $D := np$. We will suitably
  choose $\eps$ later.

  Fix a set $S \subseteq V$ of size $i$. We wish to upper bound the
  probability that $S$ makes the conductance of $G$ undesirably low,
  i.e., the probability that $E(S, \bar{S}) < \varphi \Delta(S)$.
  Denote this probability by $p_S$. By the definition of conductance
  and $(D, \eps)$-uniformity of $G$, we only need to consider subsets
  of size at most $\eta n$, for $\eta := (1+\eps)/2(1-\eps)$.

  There are $i(n-i)$ ``potential'' edges between $S$ and its
  complement in $G$, where each edge is taken independently at random
  with probability $p$. Therefore, the expected size of $E(S,
  \bar{S})$ is
\begin{displaymath}
 \nu := D i(1-i/n) \geq D i (1-\eta).
\end{displaymath}
  Now note that
  the event $E(S, \bar{S}) < \varphi \Delta(S)$ implies that
\begin{displaymath}
E(S,\bar{S}) < \varphi Di(1+\eps) < \varphi' \nu,
\end{displaymath}
   where $\varphi' :=
  \varphi(1+\eps)/(1-\eta)$.  So it suffices the upper bound the
  probability that $E(S, \bar{S}) < \varphi' \nu$.  Note that, since
  $\varphi < 1/2$, we can choose $\eps$ small enough to ensure that
  $\varphi' < 1$.  Now, by a Chernoff bound,
  \begin{align*}
  p_S &\leq \Pr[ E(S, \bar{S}) < \varphi' \nu ] \\
  &\leq \exp(-(1-\varphi')^2 \nu) \\
  &\leq  n^{-i \alpha (1-\varphi')^2
    (1-\eta)}.
\end{align*}
  Set $\alpha$ large enough (i.e., $\alpha \geq
  2/(1-\varphi')^2(1-\eta)$) so that the right hand side becomes at
  most $n^{-2i}$. Therefore, with high probability, for our particular
  choice of $S$ we have $E(S, \bar{S}) / \varphi(S) \geq \varphi$.

  Now we take a union bound on all possible choices of $S$ to upper
  bound the probability of conductance becoming small as follows.
  \begin{align*}
  \Pr[\Phi(G) < \varphi] &\leq \sum_{i=1}^{\eta n} \binom{n}{i} n^{-2i}\\
  &\leq \sum_{i=1}^{\eta n} n^{-i} = o(1).
\end{align*}
  Thus with probability $1-o(1)$, we have $\Phi(G) \geq \varphi$.

\end{IEEEproof}
By combining Lemma~\ref{lem:randomExp} and Theorem~\ref{thm:JS} we get
the following corollary, which is formally proved in Section~\ref{pr:coro:gnpMixing}.
\begin{coro} \label{coro:gnpMixing} There is an $\alpha > 0$ such that
  a random graph $G=G(n,p)$ with $p \geq \alpha \ln n/n$ has
  $\delta$-mixing time bounded by $O(\log(1/\delta))$ with probability
  $1-o(1)$.
\end{coro}

In particular, for our specific choice of $\delta := (1/2cn)^2$, the
$\delta$-mixing time of $G(n,p)$ would be $T(n) = O(\log n)$.

\subsection{Expander Graphs with Constant Spectral Gap}
Similar to Corollary~\ref{coro:gnpMixing},
we need to show that the mixing time of an expander graph
with second largest eigenvalue that is bounded away from $1$
is bounded by $O(\log n)$.

\begin{lem} \label{lem:Expander} If $G=(V,E)$ is an expander graph
  with a (normalized) second largest eigenvalue that is bounded away
  from $1$ by a constant, then $T(n) = O(\log n)$.
\end{lem}

\begin{IEEEproof}
We first recall a well known result in graph theory (cf.\ \cite{HLW06}), which states that any
regular graph with a normalized adjacency matrix whose second largest eigenvalue (in
absolute value) is bounded away from $1$  must have good expansion (i.e., $\sigma=\Omega(1)$).

Moreover, note that for regular graphs we have $\Delta(S)=D|S|$, and therefore
the two notions of conductance (Definition \ref{def:conductance}) and
expansion (Definition~\ref{def:EXPgraph}) coincide
(except a multiplicative constant).

Finally, we can applying  Theorem~\ref{thm:JS} to find the smallest $t$ which satisfies
\[
(1-\Phi(G)^2/2)^t \leq \frac{1}{(2n)^2},
\]
which implies $T(n)=O(\log n)$.
\end{IEEEproof}
We now have all the tools required for proving
Theorem~\ref{thm:num-mes}.

\begin{IEEEproof}[Proof of Theorem~\ref{thm:num-mes}]
  Follows immediately by combining Theorem~\ref{thm:disjunct},
  Proposition~\ref{prop:randomDegree},
  Corollary~\ref{coro:gnpMixing}, and Lemma~\ref{lem:Expander}.
\end{IEEEproof}

\bibliographystyle{IEEEtran} \bibliography{bibliography}
\section{Appendix}
\subsection{Proof of Proposition~\ref{lem:prob}}
\label{pr:lem:prob}
We can write
\begin{align*}
  \Pr[A \mid B] &=\frac{\sum_{i=1}^n \Pr[A | B_i] \Pr[B_i]}{\Pr[B]} \\
  &\leq \eps \cdot \frac{\sum_{i=1}^n \Pr[B_i]}{\Pr[B]}\\ &\leq \eps k.
\end{align*}
The last inequality is due to the fact that each element of the sample
space can belong to at most $k$ of the $B_i$, and thus, the summation
$\sum_{i=1}^n \Pr[B_i]$ counts the probability of each element in $B$
at most $k$ times.

\subsection{Proof of Proposition~\ref{prop:stationary}}
\label{pr:prop:stationary}
We start with a well-known result \cite[Theorem~7.13]{ref:MU}, that
a random walk on any graph graph $G$ that is not bipartite converges
to a stationary distribution $\mu$, where
\begin{displaymath}
  \mu(v)=\frac{d(v)}{2|E |}.
\end{displaymath}
Since $G$ is a $(D, c)$-uniform graph we know that $D\leq d(v)\leq cD$
and that $nD\leq 2|E|\leq ncD$.

\subsection{Proof of Proposition~\ref{lem:piV}}
\label{pr:lem:piV}
Let $t' := \lfloor t/T(n) \rfloor$, and for each $i \in \{0, \ldots,
t'\}$, $w_i := v_{iT(n)}$.  Denote by $W' := \{ w_0, \ldots, w_{t'}
\}$ a subset of $t'+1$ vertices visited by $W$.  Obviously, $\pi_v$
is at least the probability that $v \in W'$.  Thus it suffices to
lower bound the latter probability.

By the definition of mixing time, regardless of the choice of $w_0$,
the distribution of $w_1$ is $\delta$-close to the stationary
distribution $\mu$, which assigns a probability between $1/cn$ and
$c/n$ to $v$ (by Proposition~\ref{prop:stationary}).  Therefore,
$\Pr[w_1 \neq v \mid w_0] \leq 1 - 1/cn + \delta$. Similarly, $\Pr[w_2
\neq v \mid w_0, w_1] \leq 1 - 1/2cn$, and so on. Altogether, this
means that
\begin{align*}
  \Pr[w_0 \neq v, w_1 \neq v, \ldots, w_{t'} \neq v] &\leq (1 - 1/cn + \delta)^{t/T(n)}\\
  &\leq (1 - 1/2cn)^{t/T(n)} \\
  &\leq \exp(-t/(2cnT(n)))\\
  &\leq 1-\Omega(t/(cnT(n))).
\end{align*}
In the last equality we used the fact that $\exp(-x)\leq 1-x/2$ for $0\leq x \leq 1$.
Thus the complement probability is lower bounded by
$\Omega(t/(cnT(n))$. The calculation for $\pi_e$ is similar.

\subsection{Proof of Proposition~\ref{prop:toomany}}
\label{pr:prop:toomany}
For every $i = 0, \ldots, t$, define a boolean random variable $X_i
\in \{0,1\}$ such that $X_i = 1$ iff $v_i = v$. Let $X := \sum_{i=0}^t
X_i$ be the number of times that the walk visits $v$.  For every $i
\geq T(n)$, we have
\begin{align*}
\bbE[X_i] &= \Pr[v_i = v]\\
&\leq c/n + \delta\\
&\leq  2c/n,
\end{align*}
where the first inequality is due to the assumption that $i \geq T(n)$
and after the mixing time, the distribution induced on each vertex is
within $\delta$ of the stationary distribution, and the second
inequality is by the particular choice of the proximity parameter $\delta$.
 Define $X' :=
\sum_{i=T(n)}^t X_i$.  By linearity of expectation, $\bbE[X'] <
2ct/n$, and by Markov's inequality,
\begin{displaymath}
\Pr[X' \geq \alpha c^2 T(n)] <
\frac{2t}{\alpha cnT(n)}.
\end{displaymath}
By taking $\alpha$ a large constant (depending on the constant hidden in
the asymptotic estime of $\pi_v$ given by Lemma~\ref{lem:piV}),
and using Proposition~\ref{lem:piV}, we can ensure that
the bound on the probability  is at most $\pi_v/4$. Thus the probability that $X \geq k$ for $k
:= (1+\alpha c^2)T(n)$ is at most $\pi_v/4$.  Proof for the edge case
is similar.
\subsection{Proof of Proposition~\ref{prop:early}}
\label{pr:prop:early}
By the choice of $v$ (that is not a designated vertex), the walk $W$
has a chance of visiting $v$ as the initial vertex $v_0$ only
if it starts at a vertex chosen uniformly at random. Thus the probability of
visiting $v$ at the initial step is $1/n \leq 1/D$.

Now, regardless of the outcome of the initial vertex $v_0$, the
probability of visiting $v$ as the second vertex $v_1$ is at most
$1/D$, as $v_0$ has at least $D$ neighbors and one is chosen uniformly
at random.  Thus, $\Pr[v_1 = v] \leq 1/D$, and similarly, for each
$i$, $\Pr[v_i = v] \leq 1/D$. A union bound gives the claim.

\subsection{Proof of Proposition~\ref{prop:influence}}
\label{pr:prop:influence}
We can write
\begin{equation*}
  \Pr[v_i = u \mid v_j = v, \cE] = \Pr[v_j = v \mid v_i = u, \cE] \cdot \frac{\Pr[v_i = u \mid \cE]}{\Pr[v_j = v \mid \cE]}.
\end{equation*}
Now, from the definition of mixing time, we know that
\[
| \Pr[v_j = v \mid v_i = u, \cE] - \Pr[v_j = v \mid \cE] | \leq
2\delta,
\]
because regardless of the knowledge of $v_i = u$, the distribution of
$v_j$ must be $\delta$-close to the stationary distribution.
Therefore,
\begin{align*}
  | \Pr[v_i = u | v_j = v, \cE] - \Pr[v_i = u | \cE] | &\leq  2\delta / \Pr[v_j = v \mid \cE] \\
  &\leq  2\delta / (1/cn - \delta) \\ &\leq 8 \delta c n/3.
\end{align*}

\subsection{Proof of Proposition~\ref{prop:randomDegree}}
\label{pf:prop:randomDegree}
Let $\alpha := 6/\eps^2$ so that $np \geq \alpha \ln n$.  Take any
vertex $v$ of the graph. The expected degree of $v$ is $np$. As the
edges are chosen independently, by a Chernoff bound, the deviation
probability of $\deg(v)$ can be bounded as
\begin{align*}
\Pr[|\deg(v) - np| > \eps np] &\leq  2e^{-\eps^2 np/3}\\ &\leq 2 n^{-\eps^2
  \alpha/3} = 2/n^2.
\end{align*}
This upper bounds the probability by $2 n^{-2}$.  Now we can use a
union bound on the vertices of the graph to conclude that with
probability at least $1-2/n$, the degree of each vertex in the graph
is between $np(1-\eps)$ and $np(1+\eps)$.

 \subsection{Proof of Corollary~\ref{coro:gnpMixing}}
 \label{pr:coro:gnpMixing}
 Choose $\alpha$ large enough so that, by
 Proposition~\ref{prop:randomDegree} the graph becomes $(np(1-\eps),
 (1+\eps)/(1-\eps))$-uniform, for a sufficiently small $\eps$ and
 so that Lemma~\ref{lem:randomExp} can be applied to obtain $\Phi(G)
 = \Omega(1)$.  Let $\mu'$ be the distribution obtained by any random
 walk on $G$ in $t$ steps and denote by $\mu$ the stationary
 distribution of $G$.  Now Theorem~\ref{thm:JS} implies that,
 \[
 \| \mu' - \mu \|_\infty \leq (1-\Phi(G)^2/2)^t (1+\eps)/(1-\eps),
 \]
 and thus, it suffices to choose $t = O(\log(1/\delta))$ to have $\|
 \mu' - \mu \|_\infty \leq \delta$.

\begin{IEEEbiographynophoto}{Mahdi Cheraghchi}
received the B.Sc.\ degree in computer engineering from Sharif University of Technology, Tehran, Iran, in 2004 and  the M.Sc.\ and Ph.D.\ degrees in computer science from  Ecole Polytechnique F\'{e}d\'{e}rale de Lausanne (EPFL), Lausanne, Switzerland, in 2005 and 2010, respectively. Since October 2010, he has been a post-doctoral researcher at the University of Texas at Austin, TX. His research interests include the interconnections between coding theory and theoretical computer science, derandomization theory and explicit constructions.
\end{IEEEbiographynophoto}

\begin{IEEEbiographynophoto}{Amin Karbasi}
received the B.Sc.\ degree in electrical engineering in 2004 and M.Sc.\ degree in communication systems in 2007 from EPFL, Lausanne, Switzerland. Since March 2008, he has been a Ph.D. student at  Ecole Polytechnique F\'{e}d\'{e}rale de Lausanne (EPFL), Lausanne, Switzerland. He was the recipient of the ICASSP 2011 best student paper award and ACM/Sigmetrics 2010 best student paper award. His research interests include graphical models, large scale networks, compressed sensing and information theory.
\end{IEEEbiographynophoto}

\begin{IEEEbiographynophoto}{Soheil Mohajer}
received the B.Sc.\ degree in electrical engineering
from the Sharif University of Technology, Tehran, Iran, in 2004. He received
the M.Sc.\ degree in communication systems in 2005 and the Ph.D.\ degree in
2010, both from  Ecole Polytechnique F\'{e}d\'{e}rale de Lausanne (EPFL), Lausanne, Switzerland.
Since October 2010, he has been a post-doctoral researcher at Princeton University,
Princeton, NJ. His fields of interests are multi-user information theory,
network coding theory, and wireless communication.
\end{IEEEbiographynophoto}

\begin{IEEEbiographynophoto}{Venkatesh Saligrama}
received the Ph.D.\ degree from the Massachusetts
Institute of Technology (MIT), Cambridge, MA.
He is currently a faculty member with the Department of Electrical and Computer
Engineering, Boston University, Boston, MA. His research interests are
in statistical signal processing, information and control theory, and statistical
learning theory and its applications to video analytics. He has edited a book on
networked sensing, information, and control.

Dr.\ Saligrama has been an Associate Editor for IEEE Transactions on
Signal Processing and is currently serving as a member on the Signal Processing
Theory and Methods committee. He is the recipient of numerous awards
including the Presidential Early Career Award, ONR Young Investigator Award,
and the NSF Career Award.
\end{IEEEbiographynophoto}

\vfill 

\end{document}